\documentclass[a4paper,fleqn,usenatbib]{mnras}
\usepackage{color}
\usepackage{float}
\usepackage{graphicx}
\usepackage{subcaption}
\usepackage{newtxtext,newtxmath}
\usepackage{lastpage}

\usepackage[T1]{fontenc}
\usepackage{ae,aecompl}

\title[Cloudy synthetic transmission spectra of HD209458b]{Exonephology: Transmission spectra from a 3D simulated cloudy atmosphere of HD~209458b}

\author[S. Lines et al.]{
S. Lines,$^{1}\thanks{E-mail: s.lines@exeter.ac.uk}$
J. Manners,$^{1,3}$
N. J. Mayne,$^{1}$
J. Goyal,$^{1}$
A. L. Carter,$^{1}$
I. A. Boutle,$^{1,3}$\newauthor
G. K. H. Lee,$^{4,5,6}$
Ch. Helling,$^{4,5}$
B. Drummond,$^{1}$
D. M. Acreman,$^{1,2}$
and D. K. Sing$^{1}$     
\\ \\
$^{1}$Physics and Astronomy, College of Engineering, Mathematics and Physical Sciences, University of Exeter, EX4 4QL\\
$^{2}$Computer Science, College of Engineering, Mathematics and Physical Sciences, University of Exeter, EX4 4QF\\
$^{3}$Met Office, FitzRoy Road, Exeter, Devon EX1 3PB, UK\\
$^{4}$Centre for Exoplanet Science, University of St Andrews, North Haugh, St Andrews, Fife, KY16 9SS, UK\\
$^{5}$School of Physics and Astronomy, University of St Andrews, North Haugh, St Andrews, Fife, KY16 9SS, UK\\
$^{6}$Atmospheric, Oceanic $\&$ Planetary Physics, Department of Physics, University of Oxford, Oxford OX1 3PU, UK
}

\date{Accepted 17th August 2018}

\pubyear{2018}

\begin{document}
\label{firstpage}
\pagerange{\pageref{firstpage}--\pageref{lastpage}}
\maketitle

\begin{abstract}
We present high resolution transmission spectra, calculated directly from a 3D radiative-hydrodynamics simulation that includes kinetic cloud formation, for HD~209458b. We find that the high opacity of our vertically extensive cloud deck, composed of a large number density of sub-$\mu$m particles, flattens the transmission spectrum and obscures spectral features identified in observed data. We use the PandExo simulator to explore features of our HD~209458b spectrum which may be detectable with the James Webb Space Telescope (JWST). We determine that an 8 -- 12\,$\mu$m absorption feature attributed to the mixed-composition, predominantly silicate cloud particles is a viable marker for the presence of cloud. Further calculations explore, and trends are identified with, variations in cloud opacity, composition heterogeneity and artificially scaled gravitational settling on the transmission spectrum. Principally, by varying the upper extent of our cloud decks, rainout is identified to be a key process for the dynamical atmospheres of hot-Jupiters and shown to dramatically alter the resulting spectrum. Our synthetic transmission spectra, obtained from the most complete, forward atmosphere simulations to--date, allow us to explore the model's ability to conform with observations. Such comparisons can provide insight into the physical processes either missing, or requiring improvement.
\end{abstract}


\section{Introduction}
\label{sec:intro}


Clouds are expected to be ubiquitous in exoplanetary atmospheres \citep{marley13}. Transmission spectra obtained observationally from hot--Jupiters, highly--irradiated Jovian--like giant exoplanets, often contain a number of gas--phase atomic and molecular absorption features \citep[e.g.][]{charbonneau02,snellen10,sing11,birkby17}, possible small--particle condensates or photochemical `haze' that appears in spectra as non--H$_2$/He Rayleigh scattering \cite[e.g.][]{etangs08,pont08,nikolov15,kirk17} and evidence of clouds in the form of a large multi--wavelength opacity that can weaken water and other gaseous signatures \citep{deming13,sing16,iyer16}. The ability for clouds and haze to mute or mask entirely the underlying chemical composition and thermal structure of their host atmospheres, means an improved understanding of their atmospheric feedback may be critical in correctly interpreting observations.



\begin{figure*}
\centering
\includegraphics[scale=0.4]{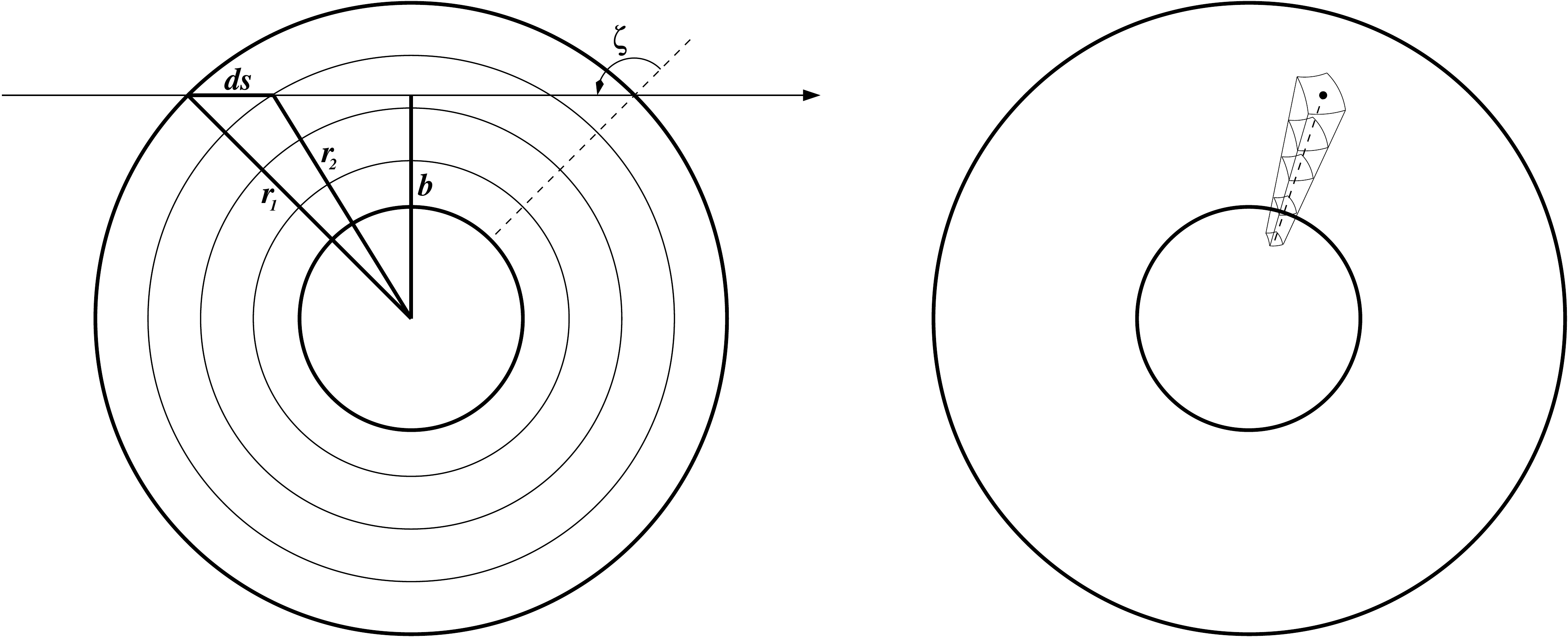}
\caption{Spherical shell geometry used for the calculation of transmission spectra. Left-hand plot shows the view perpendicular to the transit. Right-hand plot shows the view from the night side in line with the transit. Parameters are shown for a model column located in the position of the dotted line in each plot, giving a transmission spectrum at the point where the arrow leaves the top of the atmosphere (indicated by the dot in the right-hand plot). $\zeta$ denotes the stellar zenith angle, $b$ the impact parameter, and $ds$ the path length element for the layer bounded by radii $r_1$ and $r_2$. Note the path of the beam will pass through each layer twice, except for the layer in which the impact parameter is found.}
\label{fig:trans_calc}
\end{figure*}


Since cloud formation is dependent upon the local thermochemical conditions, mapping clouds is important to infer the underlying atmospheric properties. The distribution of clouds across a diverse range of planetary types is made more complex by the flow or advection, particularly from super--rotating equatorial or general zonal jets \citep[see e.g.][]{showman02,cho06,menou09,showman11,heng11,dobbs13,rauscher14,mayne14,carone15,heng15,carone16,kataria16,mayne17}, meridional advection \citep[from jet--momentum coupling, see][]{showman11,mayne13,mayne17,lines18a} and vertical mixing from a combination of mean flow (circulation) and atmospheric turbulence \citep{parmentier13}.

The swift increase in our knowledge of the dynamics and structure of hot-Jupiter atmospheres has been due to, in part, the development and adaptation of 3D atmosphere and Global Circulation Models (GCMs) which can capture both the full vertical and horizontal dynamics \citep[e.g.][]{showman02,menou09,rauscher13,mayne14}. Since the atmospheres of hot-Jupiters are heated by both a convective flux at the base of the radiative zone, and intense stellar irradiation, one important model consideration is the treatment of radiative transfer. Cloud--free simulations have produced results that closely match observations, e.g. the prediction of kilometre--per--second wind velocities from the super--rotating jets \citep{snellen10,louden15,brogi16}, and the agreement with the observed day--side emission \citep{showman09,amundsen16}. Aerosols are prevalent in the atmospheres of planets within our solar system, and the influence on their host atmospheres \citep[e.g.][]{zhang17} demonstrates clearly that neglecting the radiative feedback from clouds on their host atmospheres is not always a suitable approximation. In the last few years, a number of cloudy-GCMs, of varying complexity, have been developed \citep{parmentier13,oreshenko16,lee16,parmentier16,roman18} and advanced our understanding of cloud dynamics, radiative feedback and their effect on observables. Recently, in \citet{lines18a}, we continued this effort by coupling the Met Office GCM, the {\emph{Unified Model}} (UM) to a sophisticated kinetic, non-equilibrium cloud formation model \citep{woitke03,woitke04,helling06,helling08,lee16}. The coupled cloud-GCM model considers the homogeneous nucleation of seed particles and subsequent heterogeneous surface growth (condensation) and evaporation. The model also allows for the advection of cloud and depleted/enriched gas with the bulk atmospheric flow, gravitational settling (precipitation) and both gas and solid phase interaction with planetary and stellar radiation via absorption and scattering.


\begin{figure*}

\begin{subfigure}{0.48\textwidth}
\includegraphics[scale = 0.56, angle = 0]{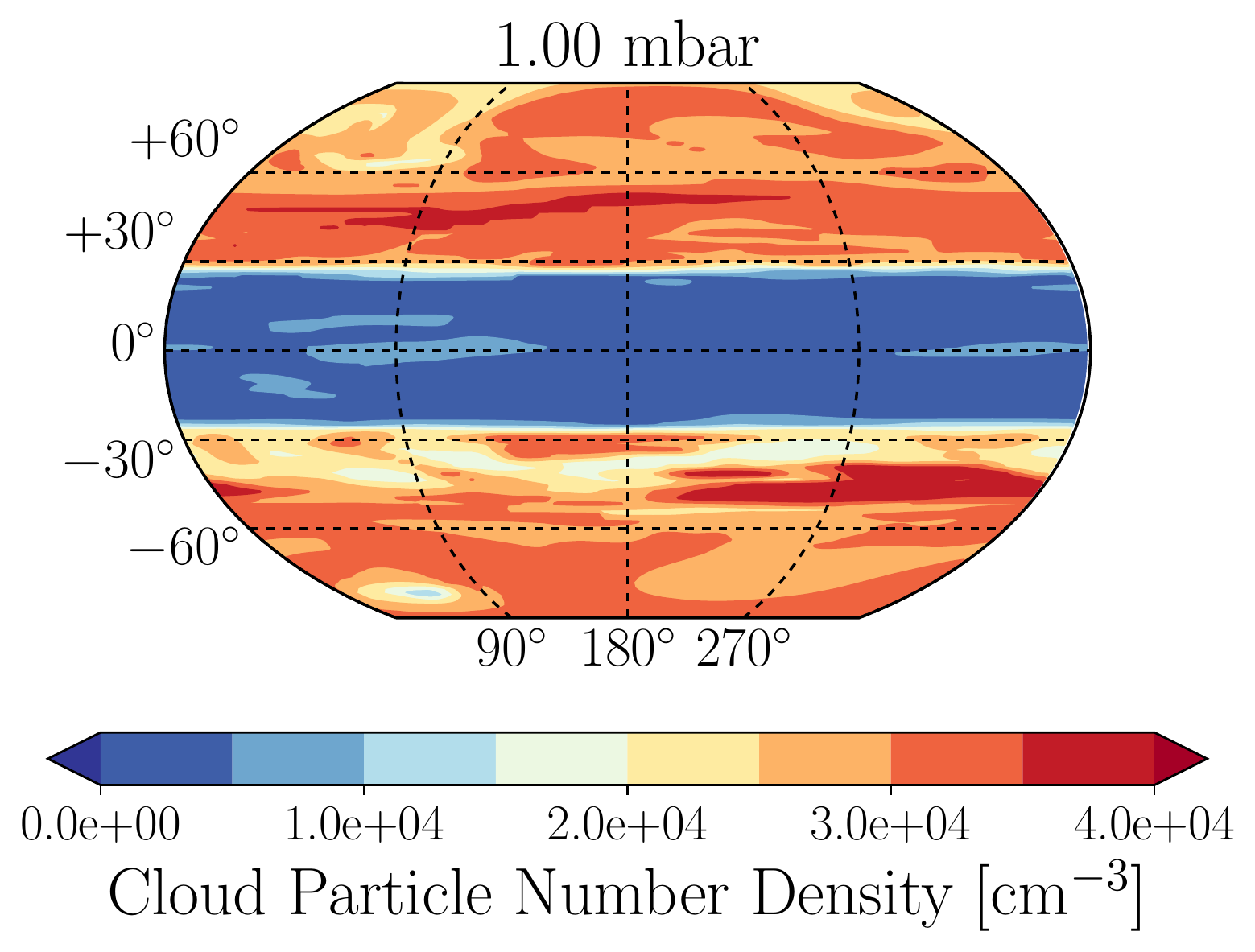}
\caption{}
\end{subfigure}\hspace*{\fill}
\begin{subfigure}{0.48\textwidth}
\includegraphics[scale = 0.56, angle = 0]{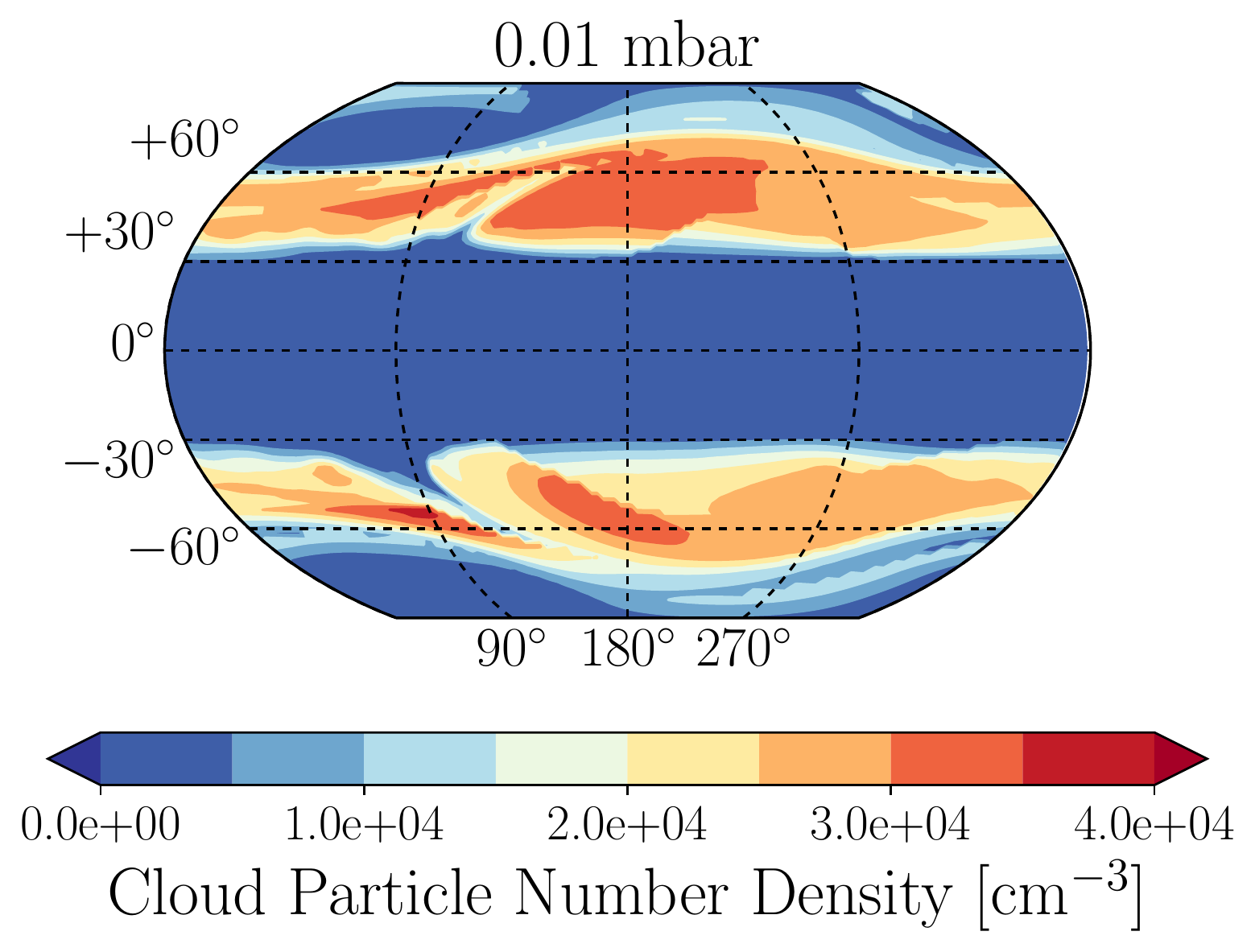}
\caption{}
\end{subfigure}

\medskip
\begin{subfigure}{0.48\textwidth}
\includegraphics[scale = 0.56, angle = 0]{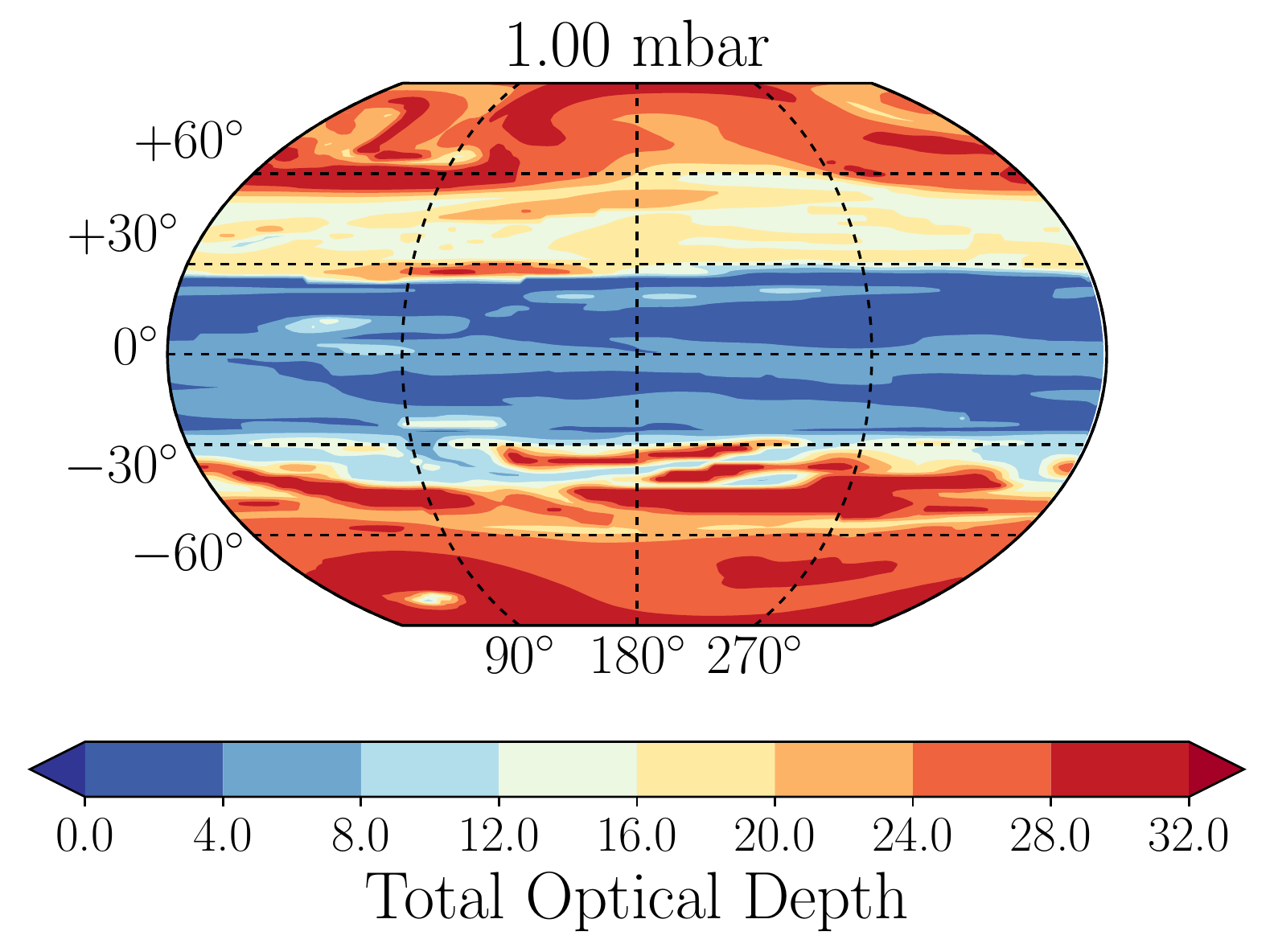}
\caption{}
\end{subfigure}\hspace*{\fill}
\begin{subfigure}{0.48\textwidth}
\includegraphics[scale = 0.56, angle = 0]{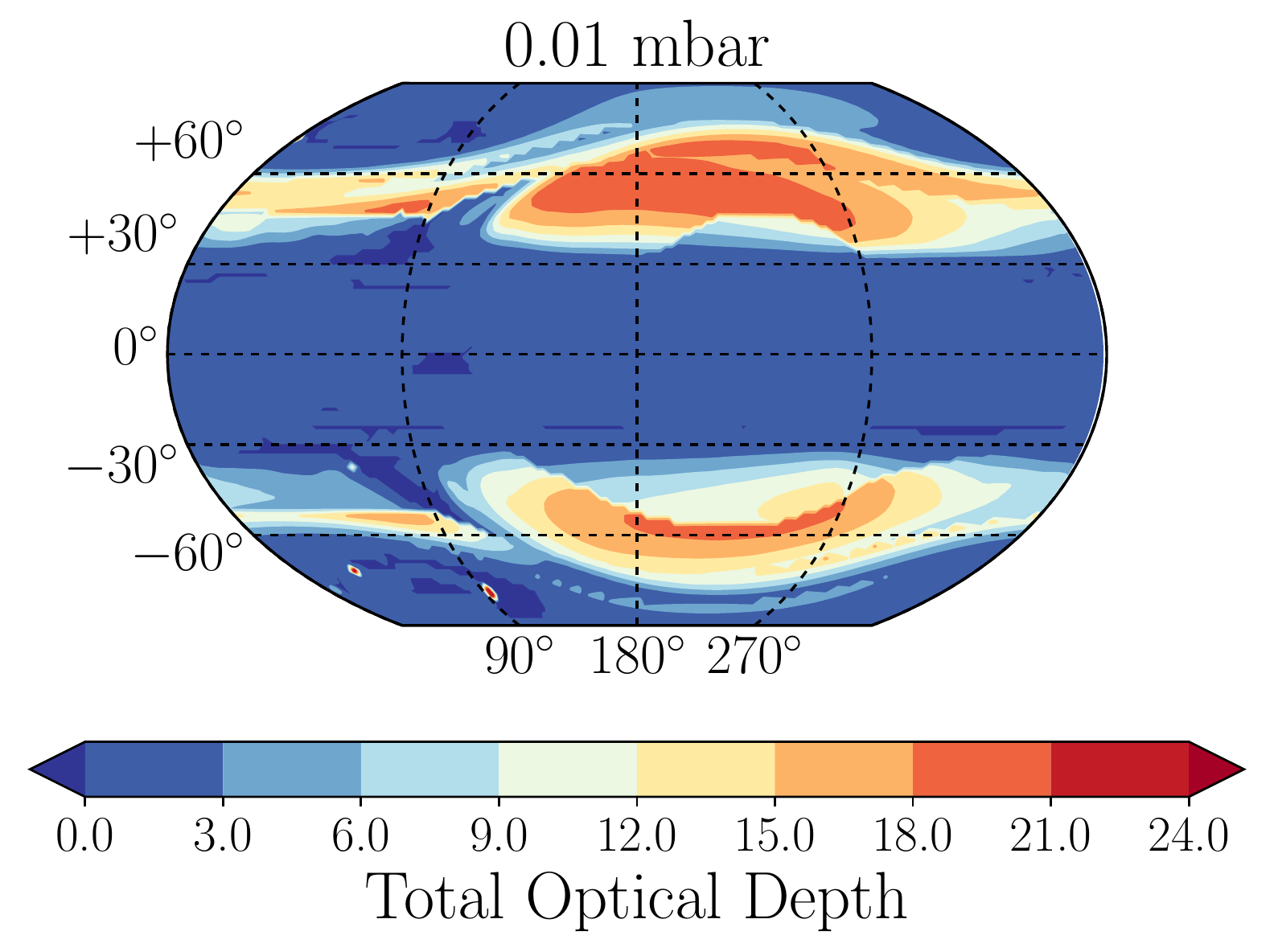}
\caption{}
\end{subfigure}

\medskip
\begin{subfigure}{0.48\textwidth}
\includegraphics[scale = 0.56, angle = 0]{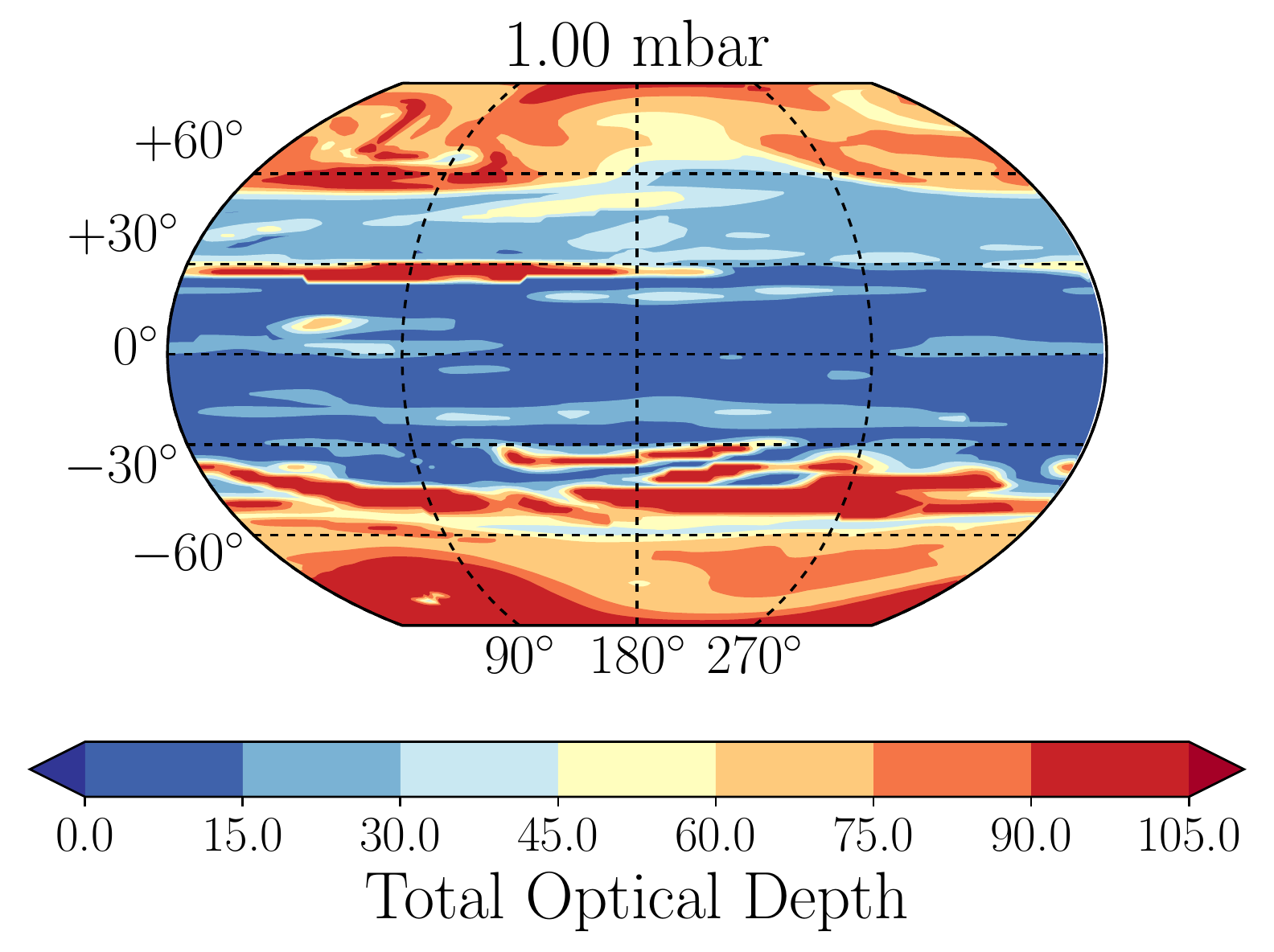}
\caption{}
\end{subfigure}\hspace*{\fill}
\begin{subfigure}{0.48\textwidth}
\includegraphics[scale = 0.56, angle = 0]{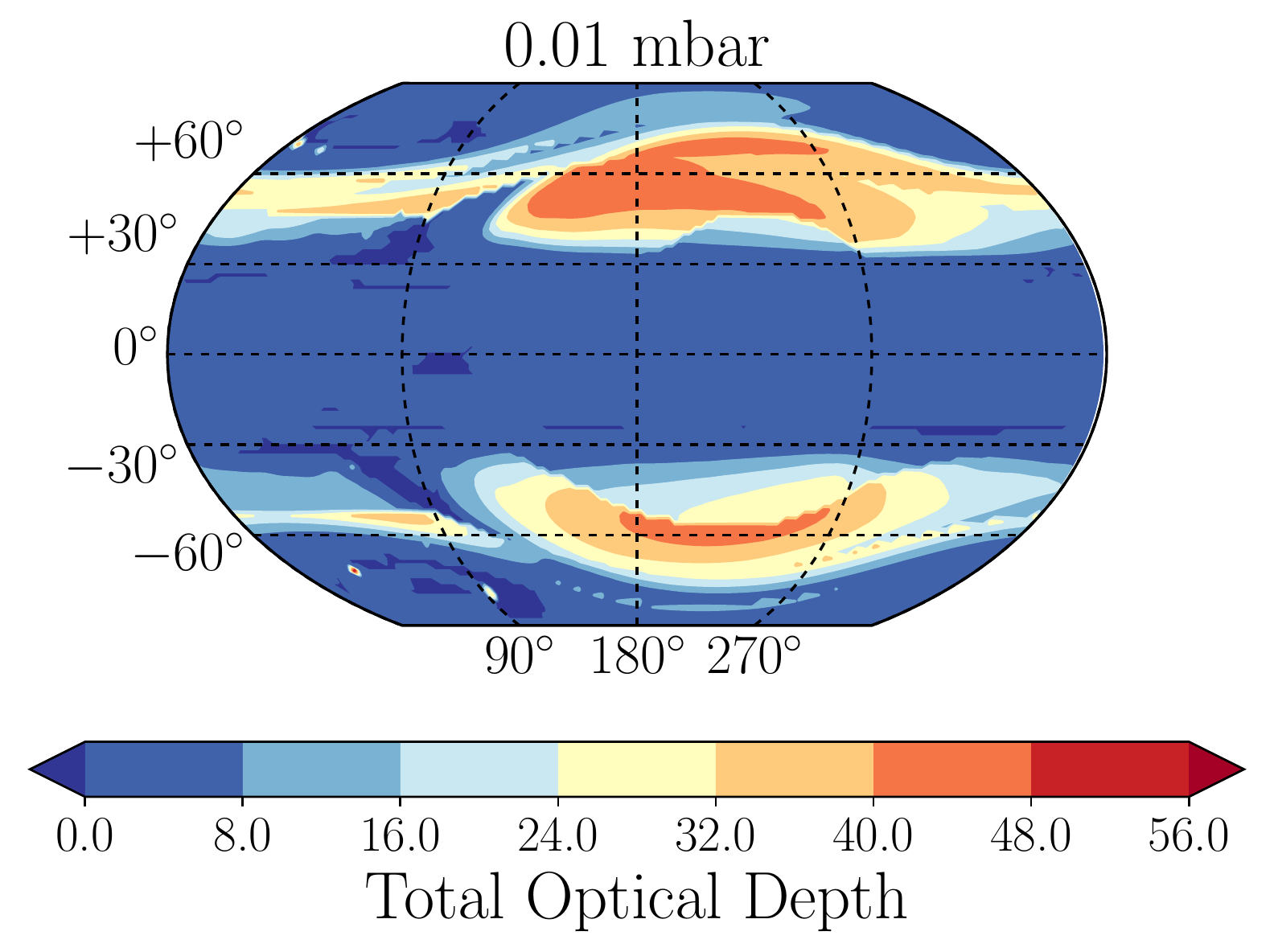}
\caption{}
\end{subfigure}

\caption{Cloud particle number density (upper), aerosol total optical depth for $\lambda$ = 6.45 - 10.40\,$\mu$m (middle) and aerosol total optical depth for $\lambda$ = 0.96 - 1.01\,$\mu$m (lower) in HD~209458b. Data obtained from the `hot' simulation at t$_{\textrm{cloud}}$ = 100 days in \citet{lines18a}. The substellar point is at $\lambda$ = 180$^o$.}

\label{fig:h209_cloud}

\end{figure*}


While theoretical modelling provides a valuable opportunity to understand atmospheric chemical processes which are not directly observable in the atmospheres of extrasolar planets, in order to verify their accuracy, comparisons to observed datasets are necessary. With respect to observations of cloud in substellar atmospheres, a combination of powerful methods have been used, including analysis of emission spectra \citep{knutson08,line16,evans17} and phase curves \citep{armstrong16,demory13,stevenson14,stevenson17}. Transmission spectra, however, remain one of the most valuable sources of atmospheric data, allowing for the direct identification of gas-phase species via their interaction with the stellar photons. This claim is supported by the wealth of transmission spectra obtained from forward models of hot-Jupiter atmospheres \citep[e.g.][]{seager00, brown01, dobbs12, showman13, wakeford15, goyal18}. With hot-Jupiters expected to contain similar chemical species, albeit in variable quantities due to their vast thermal range, the detection or absence of signatures of key species, such as the alkali metals and water vapour, can help us to infer the presence of haze and/or condensate cloud. 

One of the foremost issues with atmosphere characterisation however, is the continued absence of directly detected clouds; haze and clouds are currently indirectly inferred from non-H$_2$/He Rayleigh scattering and weakened spectral signatures from their expected broad opacity. Many studies have discussed potential condensate species operating across the JWST wavelength window \citep[e.g.][]{helling06b,helling08,min08,zeidler13,lee14,marley15,wakeford15,zeidler15,helling16,kitzmann18}. Of particular interest for mineral dust clouds forming in hot-Jupiters is the Si-O bond which has active vibrational modes operating between $\lambda$ $\sim$ 8 -- 12\,$\mu$m \citep{lee14,wakeford15}. Plotting the complex refractive index (see Figure \ref{fig:nk}) for our included dust species displays, via the extinction coefficient, the strong attenuating properties of the silicate species in the 10\,$\mu$m region can be seen.

Cloud-coupled GCMs can indicate the expected 3D distribution of cloud, precipitation efficiency and help constrain which condensate species are important for a given planet. Comparisons with observational data can only be made by producing synthetic model observables. \citet{hubbard01}, \citet{seager00} and \citet{brown01} presented some of the first studies into how model transmission spectra can be used as diagnostics to characterise giant exoplanet atmospheres. Atmospheric properties, such as the temperature and cloud cover, were investigated to reveal their impact on the resulting spectrum. While most transmission models have been analytically prescribed for 1D Pressure-Temperature (PT) profiles, \citet{fortney03} presented a 2D study and furthered this in \citet{fortney10} for application to a 3D atmosphere. Since then, various models have considered the effect of a 3D atmosphere on the transmission spectrum \citep{burrows10, dobbs12, kempton12, showman13, drummond18b, powell18}.

In this letter, we present for the first time, synthetic transmission spectra of a cloudy hot-Jupiter atmosphere which are computed directly from a state-of-the-art prognostic (predictive) and radiatively-active cloudy 3D simulation. This methodology is applied to the existing radiatively-active HD~209458b cloudy atmosphere simulation performed in \citet{lines18a} and we explore within this atmosphere, how cloud particle opacity, composition and gravitational settling affect the transmitted flux. In Section \ref{sec:theory} we introduce the details of our transmission spectrum calculation, initial conditions and methodology, in Section \ref{sec:results} we present the transmission spectra and potential for cloud detection with the James Webb Space Telescope (JWST), and in Section \ref{sec:discussion} we discuss and summarise the implications from our findings.

\section{Numerical methods}\label{sec:theory}

\subsection{Radiative Transfer}
\label{sec:rt}

For the calculation of radiative heating rates we use the open source `Suite Of Community RAdiative Transfer codes based on \citet{edwards96a}' (SOCRATES\footnote{https://code.metoffice.gov.uk/trac/socrates}) two-stream solver, in the configuration described in \citet{amundsen14}. Rayleigh scattering for our H$_2$/He atmosphere is included, and a combination of Mie and Effective Medium Theory is invoked to compute the scattering and extinction contribution from the cloud condensate particles \citep[see][for more information]{lines18a}. Scattering of both stellar and thermal fluxes is done using the Practical Improved Flux Method of \citet{zdunkowski80}. The correlated-$k$ method is used for gas absorption with absorption line data for H$_2$O, CO, CH$_4$, NH$_3$, Li, Na, K, Rb, Cs and H$_2$-H$_2$ and H$_2$-He collision induced absorption (CIA) data taken from ExoMol, and where necessary, HITRAN and HITEMP. The complete index of line list and partition function sources can be found in \citet{amundsen14}. The method of equivalent extinction  \citep[see][]{edwards96b,amundsen17} is used for the treatment of overlapping gas phase absorption.

\subsection{Transmission Spectrum Calculation}

Our transmission spectra are calculated within the 3--dimensional model framework using the same radiation scheme that is used to solve for the heating rates (i.e. SOCRATES). This can be done at any time during the simulation, by way of a second, diagnostic (not affecting model evolution), call to the scheme using a configuration with high spectral resolution. For this diagnostic call we treat the direct (unscattered) stellar radiation using spherical geometry and implement the following methodology which follows that of \citet{brown01}.

The diagnosed transmission spectrum takes into consideration the slant geometry and extinction of stellar flux through the 3D atmosphere and is made up of contributions from the direct fluxes leaving the top of the atmosphere for each GCM grid-column on the night-side of the limb. Figure~\ref{fig:trans_calc} displays the geometry considered. An important limitation of this method is that each vertical column is treated independently within the GCM, so that individual transmission spectra are derived for spherically symmetric atmospheres. For the example column displayed in Figure~\ref{fig:trans_calc} (right), the path of the direct beam in Figure~\ref{fig:trans_calc} (left) passes through model layers assuming identical optical properties to the given column, as if they were homogenous spherical shells. This calculation will then be done separately for each of the columns so that the resulting fluxes will fully represent variations across the limb perpendicular to the observer, while variations along the line of sight are only approximately represented.

Calculations are performed for model columns where the stellar zenith angle, $\zeta$, is greater than 90 degrees (i.e. lit from beneath) and where the path of the stellar beam will not intersect the bottom boundary of the model, below which the atmosphere is considered opaque.

The element of the slant path within each layer, or spherical shell, $i$ is:

\begin{eqnarray}
ds_i = & 2 \left( \sqrt{r_i^2 - b^2} - \sqrt{r_{i+1}^2 - b^2} \right) & [r_{i+1} > b], \\
ds_i = & 2\sqrt{r_i^2 - b^2} & [r_i > b, ~r_{i+1} < b], \\
ds_i = & 0 & [r_i < b].
\end{eqnarray}

The total optical depth along the slant path for a given frequency $\nu$ is then:

\begin{equation}
\tau(\nu) = \sum_{i=1}^{n} ds_i \kappa_i(\nu) \rho_i,
\end{equation}

\noindent where the sum is over the number of model layers ($n$), $\kappa_i$ is the extinction coefficient for layer $i$ and $\rho_i$ is the mean density in the layer.

The calculation is done in terms of flux in Watts per square metre normal to the direction of the incoming beam. The incident flux into the atmosphere $F_{inc}$ is determined based on the stellar constant ($S$, the stellar flux at 1AU) and the orbital distance of the planet ($D_{AU}$, in astronomical units). The outgoing flux is a simple function of the optical depth along the slant path:

\begin{eqnarray}
F_{inc}(\nu) & = & S(\nu) / D_{AU}^2,\\
F_{out}(\nu) & = & F_{inc}(\nu) e^{-\tau(\nu)}.
\end{eqnarray}

The spectral flux, $F_{out}$, here calculated as that passing through an area normal to the direction of the outgoing beam, is then converted to spectral intensity, $I_{out}$, along the beam direction by assuming the flux is spread over a solid angle equal to that subtended by the stellar disc as seen from the planet. 

This can then be used to obtain the flux seen by a distant observer of the transit which we normalise to units of Watts per square metre at 1AU:

\begin{eqnarray}
I_{out}(\nu) & = & \frac{F_{out}(\nu) D^2}{\pi R_*^2}, \\
F_{1AU}(\nu) & = & \frac{I_{out}(\nu) |dA cos{\zeta}|}{AU^2}, \\
& = & \frac{F_{out}(\nu) |dA cos{\zeta}| D_{AU}^2}{\pi R_*^2},
\end{eqnarray}

where $|dA cos{\zeta}|$ is the area of the model gridbox at the top of the atmosphere projected in the direction of the beam, $AU$ is an astronomical unit in metres, $D$ is the orbital distance in metres, and $R_*$ is the stellar radius in metres.


\begin{figure*}
\centering
\begin{subfigure}{1.0\textwidth}
\includegraphics[scale = 0.4, angle = 0]{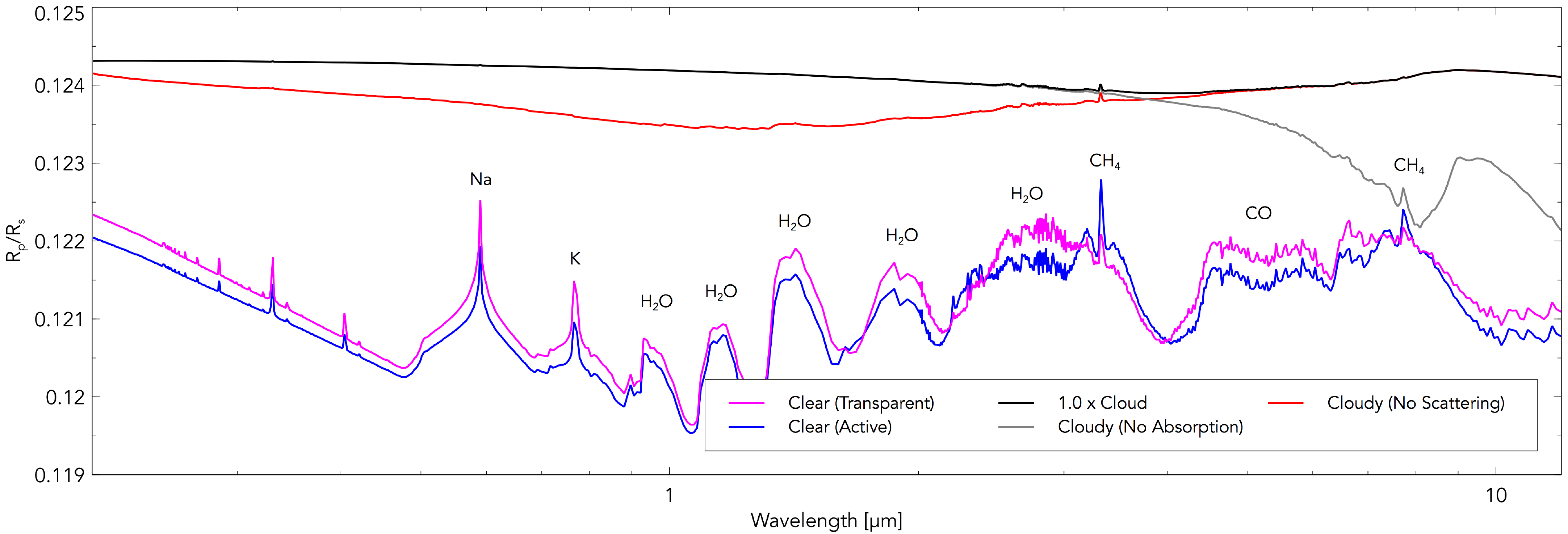}
\caption{Radiative contribution. Magenta: clear--skies atmosphere prior to cloud modification of thermal profile. Blue: clear--skies transmission from a cloudy model atmosphere (pressure--temperature profile modified by 50 days of radiatively active clouds, see \citet{lines18a} for details). Black: cloudy atmosphere including both scattering and absorption from cloud particles. Red: cloudy atmosphere without cloud particle scattering. Grey: cloudy atmosphere without cloud particle absorption.}
\end{subfigure}\hspace*{\fill}

\medskip

\begin{subfigure}{1.0\textwidth}
\includegraphics[scale = 0.4, angle = 0]{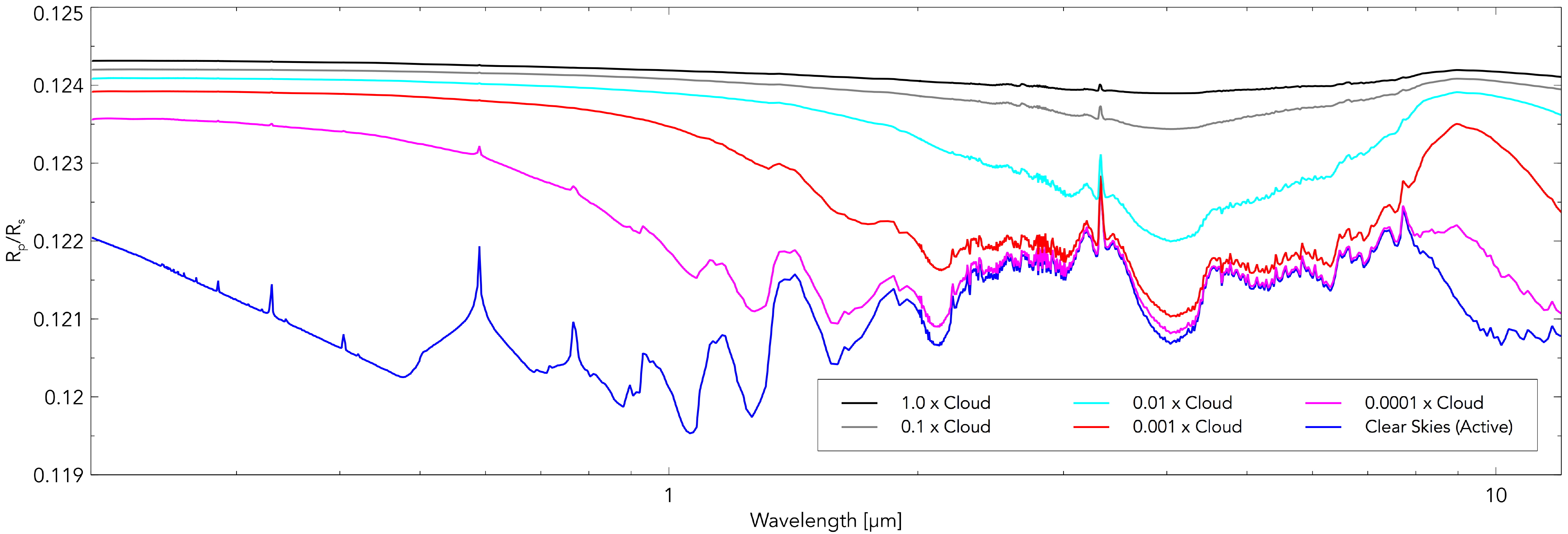}
\caption{Scaled cloud opacity. Blue: As above. Black, Grey, Cyan, Red $\&$ Magenta: cloudy atmosphere with varying scaling factor of the cloud opacity, ranging from 1.0x to 0.0001x.}
\end{subfigure}
\caption{Synthetic transmission spectra of HD~209458b.} 
\label{fig:h209_tran}
\end{figure*}


The fluxes from each contributing model column are summed to give the total flux transmitted through the planet's limb that would be received by a distant observer. We then convert this to an effective planetary radius, $R_p(\nu)$, by determining the radius of a completely opaque planet that would obscure the same amount of stellar flux. The stellar flux (normalised for an observer at a distance of 1 AU) that would be received during a transit ($S_T$) if the planet's atmosphere was totally opaque is:

\begin{equation}
S_T(\nu) = S(\nu)\left(1 - \frac{R_{p,TOA}^2}{R_*^2}\right)
\end{equation}

The actual flux observed during the transit is:

\begin{eqnarray}
S_T(\nu) & = & S(\nu)\left(1 - \frac{R_{p,TOA}^2}{R_*^2}\right) + \sum_{x,y}{F_{1AU}(\nu)} \\
& = & S(\nu)\left(1 - \frac{R_p(\nu)^2}{R_*^2}\right)
\end{eqnarray}

Then the effective planetary radius for this frequency as a fraction of the stellar radius can be found:

\begin{equation}
\frac{R_p(\nu)}{R_*} = \sqrt{\frac{R_{p,TOA}^2}{R_*^2} - \frac{\sum_{x,y}{F_{1AU}(\nu)}}{S(\nu)}}.
\end{equation}

The method employed here is somewhat different to that used by \citet{fortney10} who define a transit radius at the point where the total slant optical depth reaches 0.56. The flux based method used here, as noted by \citet{brown01}, derives directly from how the quantity is actually observed. The other major difference with the method of \citet{fortney10} is how the spherical grid is defined. \citet{fortney10} resample the output of their 3--dimensional simulations onto a grid with its pole directed towards the star. The transmission spectra for each azimuthal angle around the limb can then be calculated using the optical properties sampled along the slant path. In contrast, we calculate transmission spectra directly from the GCM as it is running using the latitude-longitude grid of the GCM without interpolation of optical properties. The advantage of this method is that transmission spectra can be diagnosed directly from the model using the full 3--dimensional information available to the radiation scheme. The treatment of each column separately within the scheme means that only the atmospheric conditions in the columns on the night side of the planet limb will be included in the calculation. This may cause biases if the optical properties of the atmosphere display strong gradients across the terminator. We address the bias introduced by our approximation in the results.

\subsection{Initial Conditions}
\label{sec:ic}

\begin {table}
\begin{center}
\begin{tabular}{ l|c }
{\bf{Parameter}} & {\bf{Value}}\\
\hline
Horizontal resolution (Grid Cells) & $\lambda$ = 144, $\phi$ = 90 \\ 
Vertical resolution (levels) & 66 \\ 
Hydrodynamical timestep (s) & 30 \\ 
Radiative timestep (s) & 150 \\ 
Intrinsic temperature (K) & 100 \\
Initial inner boundary pressure (Pa) & 2.0 x 10$^7$ \\
Upper boundary height (m) & 1.0 x 10$^7$ \\
Ideal gas constant, $R$ (Jkg$^{-1}$K$^{-1}$) & 3556.8 \\
Specific heat capacity, $c_{\textrm{p}}$ (Jkg$^{-1}$K$^{-1}$) & 1.3 x 10$^4$ \\
Radius, $R_{\textrm{p}}$ (m) & 9.00 x 10$^7$ \\
Rotation rate, $\Omega$ (s$^{-1}$) & 2.06 x 10$^{-5}$ \\
Surface gravity, $g_{\textrm{p}}$ (ms$^{-2}$) & 10.79 \\
Semi-major axis, $a_{\textrm{p}}$ (au) & 4.75 x 10$^{-2}$ \\
\end{tabular}
\vspace{+10pt}
\caption{Selected model parameters from our hot interior HD~209458b atmosphere, covering grid setup, run-lengths and planet constants. See \citet{lines18a} for more information.}
\label{tab:params}
\end{center}
\end{table}

We apply our transmission model to the final output from our `hot-interior'\footnote{Atmosphere initialised with a high entropy interior, as per the heated deep atmosphere of \citet{showman02} and \citet{tremblin17}). See \citet{lines18a} for more details.}, radiatively-active cloudy simulation of HD~209458b in \citet{lines18a}. In that work we evolved from rest, using the UM, a {\emph{cloud-free}} atmosphere of HD~209458b for t = 800 days\footnote{All references to days refer to one earth day.} followed by t$_{\textrm{cloud}}$ = 100 days of cloud formation and evolution (of which, in the final 50 days, the clouds are allowed to radiatively feedback, via absorption and scattering, onto the atmosphere). The kinetic, microphysical \citet{helling06} model is used to compute TiO$_2$ seed particle nucleation and the growth and evaporation of cloud particles, of which a single particle can be a mixture of our included condensation species (TiO$_2$, SiO, SiO$_2$, MgSiO$_3$ and Mg$_2$SiO$_4$). Cloud particles are advected and precipitated through the atmosphere, in addition to contributing to the local heating rates via the scattering and absorption of thermal and stellar irradiation. The sub-grid particle size distribution is unimodal, meaning that each cell reports a mean cloud particle size that varies between cells. The grid setup and planet constants can be found in Table \ref{tab:params} and a brief overview of the cloud distribution and radiative properties is displayed in Figure \ref{fig:h209_cloud}.

\subsection{Methodology}

Since our transmission method is performed at `run--time', (i.e. at any time requested during the execution of the UM), it is only necessary to run our simulations for a single t = 30s hydrodynamical time--step in order to retrieve the transmitted flux. From \citet{lines18a} and described in Section \ref{sec:ic}, we took the existing `hot' HD~209458b cloudy atmosphere, at t = 100 earth days of simulated cloud formation and continued this simulation under the same conditions for a single hydrodynamical (and radiative) time--step. During this step, a second {\emph{diagnostic}} call to the radiative transfer scheme was enabled whereby the transmitted flux is calculated using spectral files that cover $\lambda$ = 0.2 -- 10,000\,$\mu$m using 950 bands, a significant improvement over the 32 bands used to obtain the heating rates \citep[see][for more information]{amundsen14}.

We simulated, and extracted the transmitted flux from, atmospheres corresponding to 18 scenarios that explore the roles of cloud opacity, composition and gravitational settling. In the first study, we analysed two clear--sky atmospheres with pressure--temperature (PT) profiles corresponding to an atmosphere prior to and after modification by radiatively active clouds, a cloudy simulation with full opacity from our mixed composition cloud particles, including both scattering and absorption by cloud particles, one without cloud particle absorption and one without cloud particle scattering. In the second study we added a further four cases in which we scaled the cloud opacity (by way of the scattering and absorption coefficients) by 0.1, 0.01, 0.001 and 0.0001.

In the third study we investigated the role of composition heterogeneity on the transmission spectrum by forcing mixed composition cloud particles to adopt a single chemical make-up (as if cloud has formed by homogenous condensation) using each of the five contributing dust species that make up our original cloud particles; TiO$_2$, SiO, SiO$_2$, MgSiO$_3$ and Mg$_2$SiO$_4$. This process involved setting the volume contribution to 100$\%$ for each species, in turn, and thus bypassing the need to calculate average optical values via effective medium theory.


\begin{figure*}
\centering
\includegraphics[scale=0.4]{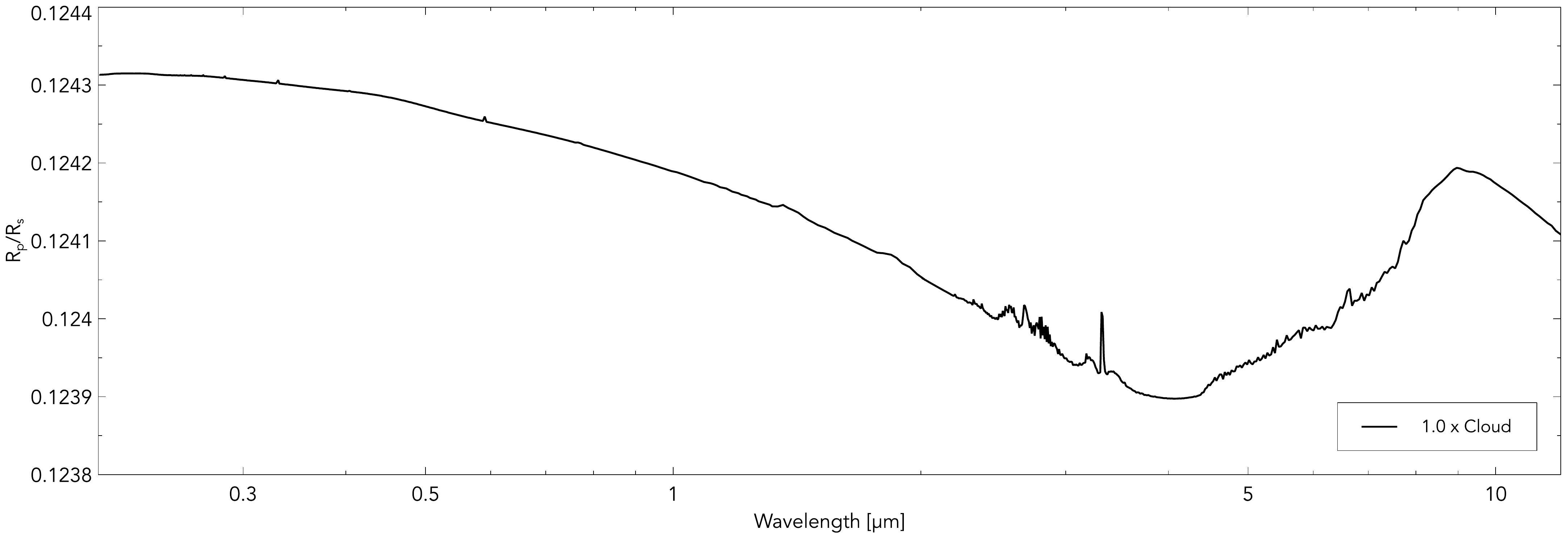}
\caption{Simulated transmission spectrum from the full cloud opacity model. As per Figure \ref{fig:h209_tran} but with R$_p$/R$_s$ axis scaling to show the spectrum shape.}
\label{fig:tran_zoom}
\end{figure*}


In our final study, we explored the role of cloud particle precipitation. Gravitational settling of cloud may significantly alter the atmospheric observables as it can be an efficient way to remove high opacity condensate particles and haze from the upper atmosphere. Precipitation timescales can be long; from \citet{lines18a} we find that particles settling, for example from P = 1 - 2 bar, can occur within t = 50 days providing there is no other advective process at play. However, this gravitational rain-out can be (and in our case is) off-set by upwards vertical transport from atmospheric circulation in the form of global winds. This reduction in the net cloud particle velocities (10$^{-1}$ to 10$^{-4}$ ms$^{-1}$) results in a required integration time to a semi-steady state of tens of years. In some cases, strong day-side up-draughts can entirely support or even loft small particulates to high altitudes. The equilibrium state of such a process is therefore not currently possible to obtain with computationally demanding 3D GCMs. Therefore, to simulate the effects of an atmosphere in which cloud has settled to deeper pressures, we executed our transmission scheme for four additional cloudy profiles whereby the cloud opacity is zeroed for the first (upper) 15, 20, 25 and 30 vertical layers, corresponding to upper cloud boundaries of approximately P = 0.01, 0.1, 1 and 10 mbar. Since pressure, for a given height, varies as a function of horizontal location, the cloud top pressure varies slightly for a single layer\footnote{The UM grid uses a vertical grid on geometric height instead of pressure.}.

\section{Results}\label{sec:results}

\subsection{Clear Skies}

Our synthetic transmission spectra are displayed in Figure \ref{fig:h209_tran}. With the clear--sky (radiatively transparent clouds) simulation we see H/He Rayleigh scattering up to 0.5\,$\mu$m that is immediately followed by the prominent Na and K absorption lines, and at longer wavelengths by H$_2$O and CO features. The clear transmission (active model clouds) simulation considers an atmosphere that does not contain direct scattering or absorption from cloud particles, but has a PT profile that has adjusted in response to 50 days of radiatively active clouds. By comparison with the `transparent' clear-skies simulation (which considers a PT profile prior to radiatively active clouds), the strongly scattering and hence cooler atmosphere (up to 250 K) results in a modification to the solution of gas--phase chemical equilibrium. This effect reduces the alkali metal signatures and, for $\lambda$ $<$ 3\,$\mu$m, water features. Most noticeable though is the CH$_4$ feature at 3.3\,$\mu$m (and to a lesser extent 7.8\,$\mu$m), which has an enhanced absorption amplitude; a result of the cooler conditions increasing the methane abundance (in relation to CO).

\subsection{Cloudy Skies}

The high cloud particle number densities introduce a large cloud opacity that spans across both the visual and infra--red leading to a flat spectrum compared to a clear-skies atmosphere. However, when isolating the full cloud opacity model as shown in Figure \ref{fig:tran_zoom}, the transit radius ratio is revealed to have a wavelength dependence. Notably, the extensive Mie scattering from our 0.3 $\mu$m cloud particles leads to a shallower gradient in the visual, reducing the H/He Rayleigh scattering which is a prominent feature a clear-sky atmosphere. Beyond 4\,$\mu$m, cloud particle absorption begins to play a stronger role and leads to a cloud feature which is extremely broad (due to both a mixed composition and particle size) but peaking at around 9\,$\mu$m. The only gas-phase feature that is not completely flattened by the broad cloud opacity is the narrow CH$_4$ signature at 3.3\,$\mu$m.

In the scatteringly--only simulation, the transmitted flux increases (R$_\textrm{p}$/R$_\textrm{s}$ decreases) for wavelengths longer than that of the methane feature (3.3\,$\mu$m). Interestingly, cloud scattering also produces a broad feature between 8 -- 12\,$\mu$m, although at these wavelengths the cloud extinction is entirely dominated by absorption. The absorption--only simulation produces almost identical spectra to the full simulation for $\lambda$ $>$ 4.5\,$\mu$m, indicating the 9\,$\mu$m peak is attributed to cloud absorption.

Decreasing the cloud opacity helps to accentuate the 9\,$\mu$m cloud feature, as it reduces the opacity outside of this window, although the strength of the features attenuates for the lowest opacity (0.0001 x). The 0.001 x scaling factor leads to a well defined cloud absorption feature as well as beginning to reveal the water vapour and carbon monoxide features. Since the Mie scattering efficiency is high around our average cloud particle size of 0.3\,$\mu$m \citep[see][for more information]{lines18a}, the opacity remains high in the optical even for the highest opacity scaling. As a result, the alkali metal features are still heavily muted.

To address and quantify the bias in our transmission model, which considers only the atmospheric properties from the nightside columns, we perform an additional simulation that is not shown in Figure \ref{fig:h209_tran}. A spectrum is obtained for the full cloud opacity (1.0 x cloud) model, but we instead perform the transmission calculation taking into account only the properties from the dayside. This spectrum is then averaged with the nightside only model to correspond to a solution which accounts for the atmospheric properties across both sides of the terminator. We find an offset (increase) in the transit depth of between 25 - 50 ppm, which is unlikely to be detectable through the instrument noise. Such an increase in the atmospheric opacity is expected, due to higher cloud particle number densities on the {\emph{dayside}}, leading to increased optical depths; these cloud properties are illustrated in Figure \ref{fig:h209_cloud}. Transmission spectra obtained observationally have a floating baseline radius which is unknown and therefore must be deduced by fitting routines. The resulting spectrum is therefore not on an absolute radius (or pressure) scale, making the identification of {\emph{relative}} differences in the spectral features a critical aspect in characterisation. In our case, the overall shape of the spectrum remains consistent between models using the nightside and the averaged values. Crucially this allows for the same identification of spectral markers, such as the silicate peak, in addition to resolving overall trends in the data.


\begin{figure*}
\centering
\includegraphics[scale=0.4]{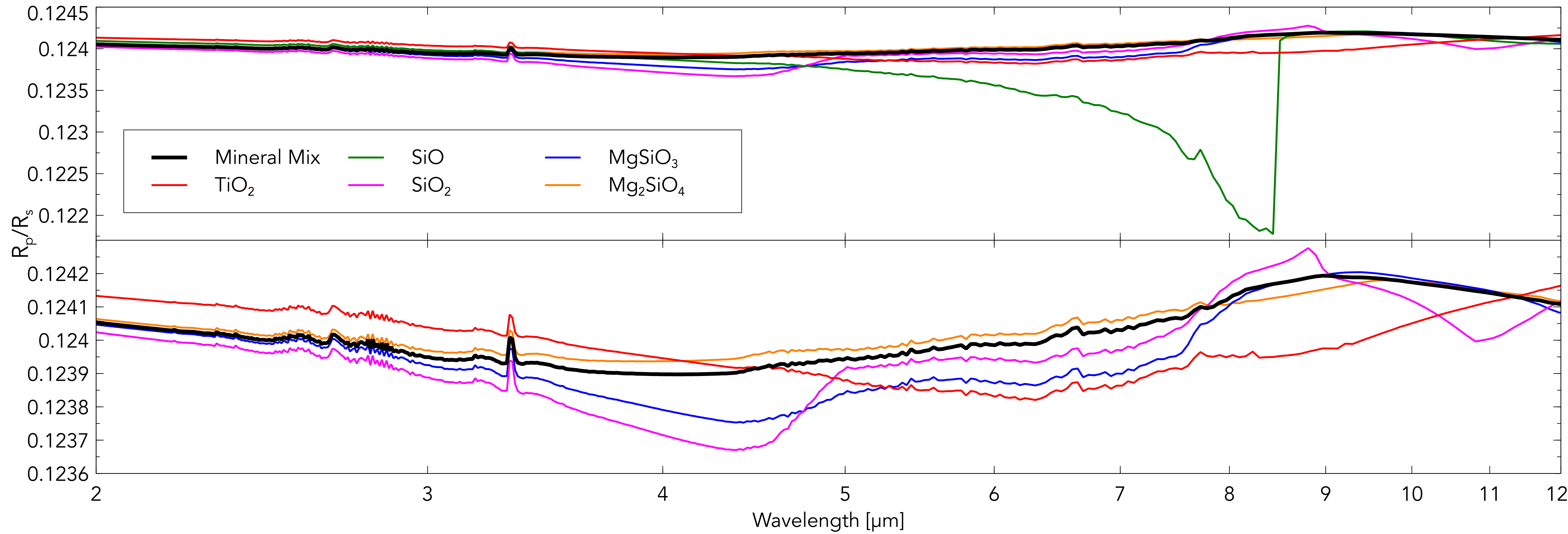}
\caption{Synthetic transmission spectrum of our cloudy HD~209458b. Particle composition is forced from variable mineral mixture (mixed composition with volume ratios taken from \citet{lines18a}) to single species. The same data, without SiO, is shown in the lower plot.}
\label{fig:composition}
\end{figure*}


\subsection{Composition}

In Figure \ref{fig:composition}, we present the transmission spectra for our forced cloud composition. While forcing a purely SiO cloud composition presents a dramatically different transmission profile for 5\,$\mu$m $<$ $\lambda$ $<$ 8.5\,$\mu$m the remaining cloud species closely follow the default mineral mix. Pure TiO$_2$ introduces a higher opacity in the visual and near-IR, but does not produce the cloud absorption features (attributed to the Si-O bonding in the silicates) at 9\,$\mu$m. TiO$_2$ regains dominance from 12\,$\mu$m due to the its high extinction factor at 20\,$\mu$m. However, in our upper atmosphere TiO$_2$, alongside SiO, contributes very little (or in some cases none) of the cloud particle volume; contributions from TiO$_2$ exist primarily in the form of the nucleation seed, and that from SiO is only non-negligible for the deepest atmosphere \citep[see][for more information on cloud particle composition]{lines18a}. At pressures probed by transmission spectroscopy, the compositional mix is approximately 15$\%$ SiO$_2$, 30$\%$ MgSiO$_3$ and 55$\%$ Mg$_{2}$SiO$_4$.

While the opacities from the magnesium silicates peak at similar wavelengths (9-10\,$\mu$m), SiO$_2$ provides an opacity maximum, for the wavelengths plotted in Figure \ref{fig:composition}, at shorter values which helps to broaden the mineral mix absorption peak towards 8\,$\mu$m.

\subsection{Precipitation}

\begin{figure*}
\includegraphics[scale=0.4]{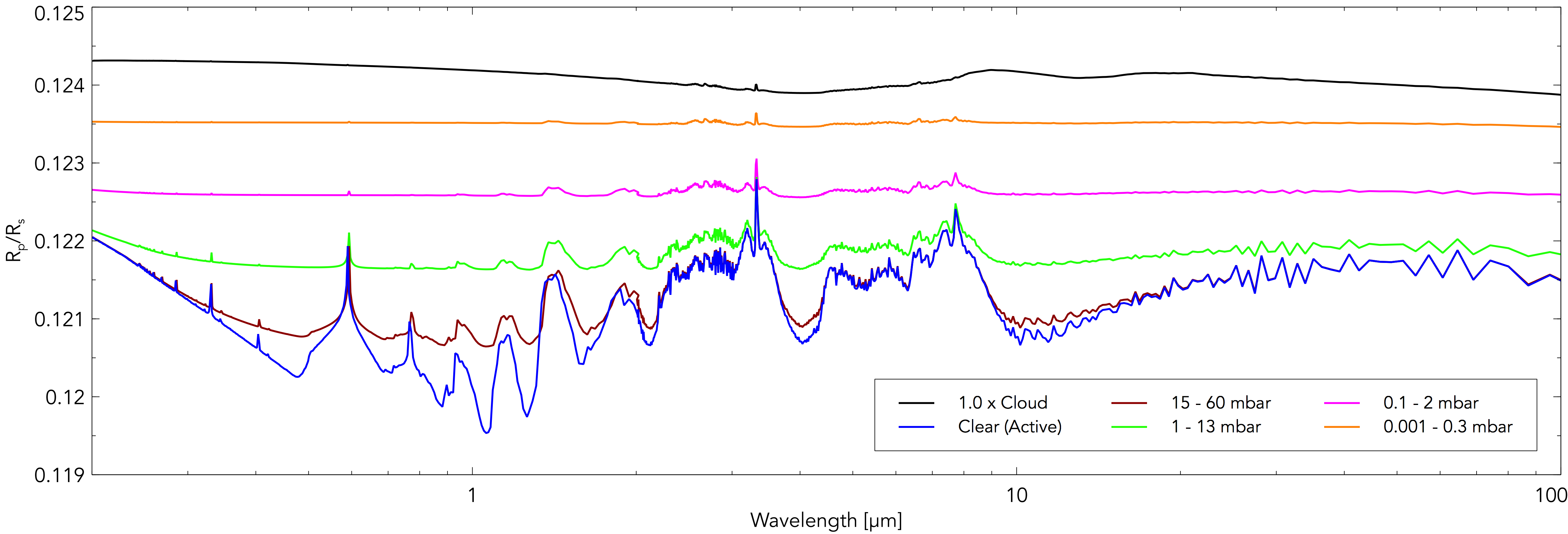}
\caption{Transmission spectra exploring the effects of gravitational settling of cloud particles, produced from the set of `enhanced precipitation' simulations. Pressures in the legend indicate the minimum and maximum pressures traced by the cloud top (upper cloud boundary) in each model.}
\label{fig:tran_settle}
\end{figure*}

In Figure \ref{fig:tran_settle}, we present the transmission spectra of our cloudy atmosphere with enhanced precipitation, calculated by zeroing the cloudy opacity in the upper $n$ layers. By removing cloud scattering and absorption in the first 15, 20, 25 and 30 vertical layers, we force the cloud top to pressures approximately P = 0.01, 0.1, 1 and 10 mbar respectively, simulating the effects of cloud particle rainout over timescales longer than our 3D GCM can practically capture. The results indicate that the cloud must settle to the millibar pressure level until all spectral features common to HD~209458b (Na, K and H$_2$O) can be seen. Cloud confined to below P = 1 mbar results in a grey opacity, with a weak sodium signal and water bands. At the shortest wavelengths, the Rayleigh scattering from the background H/He atmosphere becomes important. For the deepest cloud top at P = 15 - 60 mbar, the spectrum closely traces our clear-skies atmosphere for $\lambda$ $<$ 0.3\,$\mu$m and $\lambda$ $>$ 2\,$\mu$m.

\subsection{JWST Potential}

\begin{figure*}
\includegraphics[scale=0.4]{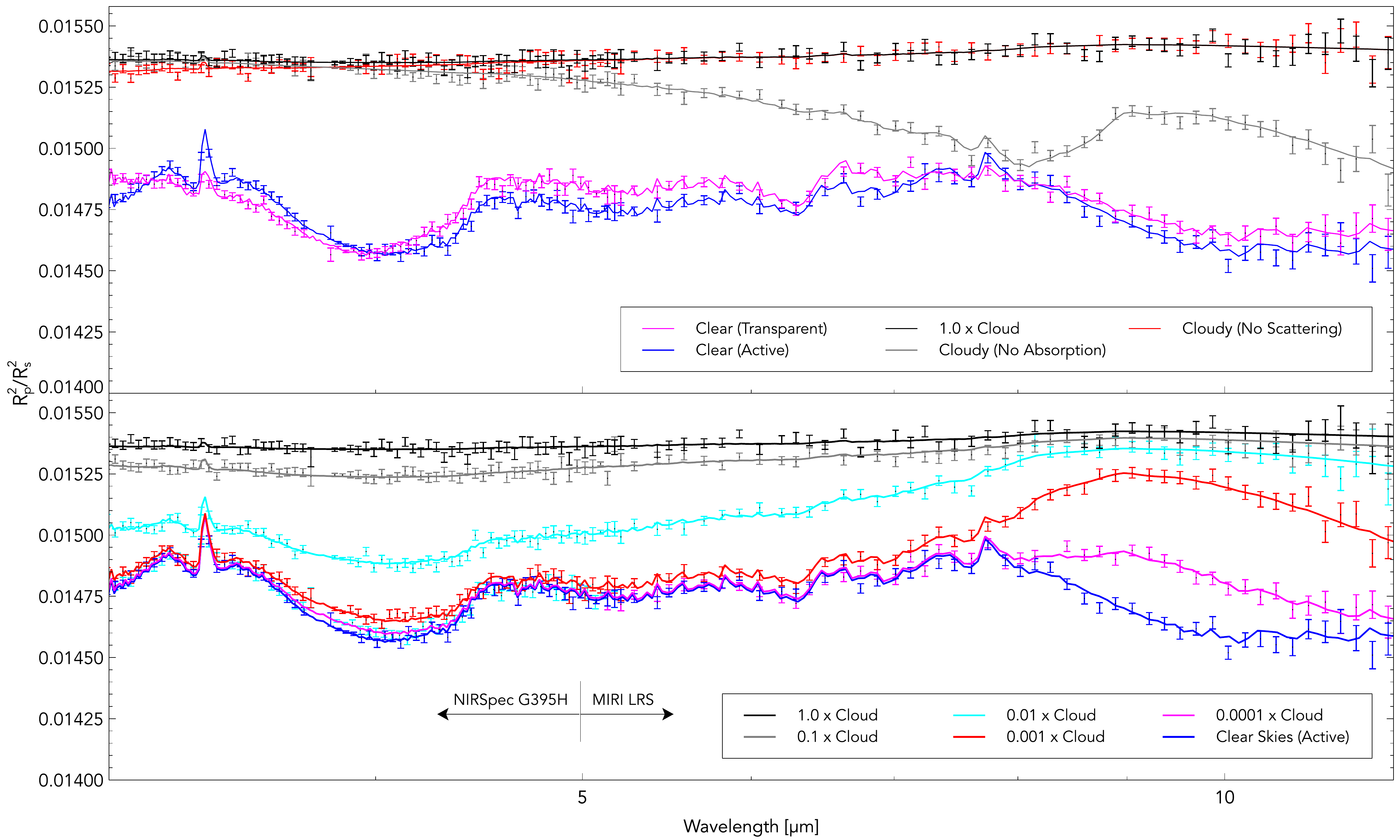}
\caption{JWST detectability with PandExo simulator showing each simulation from Figure \ref{fig:h209_tran} with the same format and colour scheme.} 
\label{fig:h209_jwst}
\end{figure*}

In preparation for the launch of JWST, \citet{batalha17} have developed a noise simulator, called PandExo, which generates simulated observations of all observatory--supported time--series spectroscopy modes. In Figure \ref{fig:h209_jwst}, we present PandExo simulations for a selection of the generated transmission spectra of HD~209458b using both the NIRSpec G395H and MIRI LRS modes.

The simulations were performed for a single occultation with an equal fraction of in, to out of, transit observation time; a noise floor of 50 ppm was set for all observation modes and detector saturation was set at 80$\%$ full well. The stellar and planetary parameters necessary for the simulation were retrieved from the TEPCAT database and the stellar spectrum used was identical to the one used in the GCM. All instrument related parameters, such as subarrays and readout patterns, were kept at the PandExo defaults. The resolution of the generated spectra are not strictly as high as the achievable resolution of the NIRSpec G395H or the MIRI LRS. However binning of the data will be typically necessary to improve the signal to noise and make resolving certain spectral features possible. As such, we have binned the NIRSpec G395H and MIRI LRS data to a resolution of R$\sim$60 and R$\sim$30 respectively. 

We measure the silicate feature between 4 and 9 $\mu$m to be 70 ppm and thus, providing the MIRI instrument noise floor lies well below the feature size, is in a regime where detection is possible. We perform a $\chi^{2}$ analysis between the PandExo simulated observations and a) our simulated full opacity cloudy model ($\chi^{2}_{\textrm{cloudy}}$) and b) a flat line `grey' spectrum ($\chi^{2}_{\textrm{grey}}$) with variable transit depth between 0.0153 $\le$ (R$_p$/R$_s$)$^2$ $\le$ 0.0155. We find $\chi^{2}_{\textrm{cloudy}}$ = 117 and 228 $\ge$ $\chi^{2}_{\textrm{grey}}$ $\le$ 12030 for 122 degrees of freedom, and rule out a fully grey atmosphere at 5.6 $\sigma$. For all cases $\chi^{2}_{\textrm{cloudy}}$ $<$ $\chi^{2}_{\textrm{grey}}$, indicating detection of the silicate feature over that of a fully grey atmosphere is favourable. The CH$_4$ amplitude of 20 ppm in the full cloud model is within the scatter and therefore is unlikely to be detected above the noise. The both the silicate and methane detectability increases, however, with reducing cloud opacity. This means that for an atmosphere with less cloud particles in the transit region (potentially originating from, for example, lower metallicity and/or enhanced settling) the detection likelihood of cloud particles, and gas absorption, improves. An opacity scaling of 0.01 x is likely required in order to detect the 3.3\,$\mu$m methane feature (150 ppm).


\section{Discussion $\&$ Summary}\label{sec:discussion}

Since our simulations produce vertically extensive cloud condensate decks with large cloud particle number densities ($>$ 3 x 10$^4$ cm$^{-3}$) even at the lowest pressures, it is not surprising that the blanket opacity across the visual and infra--red mask, almost completely, the atomic and molecular signatures in the gas--phase. Despite this, the obtained synthetic transmission spectrum for the full cloud opacity model still contains potentially detectable features. Firstly there is the flat nature, compared to a clear-skies atmosphere, of the spectrum. This indicates the presence of high opacity aerosols that can efficiently obscure the spectral signatures of the majority of gas phase absorbers, unless the cloud opacity is scaled down or the cloud deck upper boundary or `cloud top' is forced to higher (deeper) pressures. Secondly, there remains a tiny signature from CH$_4$ absorption which is amplified from the cooler atmospheric conditions produced by highly scattering cloud in our \citet{lines18a} simulations. Finally, the 9\,$\mu$m cloud absorption feature, shown to be broadened by mixed-composition cloud particles, is potentially detectable by JWST, providing a reduced cloud opacity. The previously predicted, Si-O vibrational absorption from cloud particles \citep{lee14,wakeford15}, leads to strong absorption (shown via the imaginary component of the material refractive indices) at around 10\,$\mu$m for the SiO$_2$, TiO$_2$, MgSiO$_3$ and Mg$_2$SiO$_4$ condensates considered in our work. Further understanding of how this feature changes in response to the cloud composition will likely prove critical in determining cloud properties (e.g. composition, particle/droplet size) of observed atmospheres.

We address the bias introduced by an approximation in our transmission model, and find that the offset in the transit radius ratio when considering atmospheric properties about both the night and daysides of the terminator, is minimal. The offset, which for our case lies below the JWST instrument noise, is a function of the cloud longitudinal asymmetry level. To address planetary atmospheres which could feature stronger asymmetries in the cloud distribution, future work may need to improve upon our approximation.

One aspect of cloud formation and evolution that could exert a strong influence on the cloud vertical structure and observable properties, is the gravitational settling of cloud and/or haze particles. While cloud particles can obtain large precipitation velocities \citep[see][]{lines18a}, this fall-speed can be offset by a combination of vertical mean winds that result in a slow net downwards advection of cloud. Cloud particles may also be transported by turbulence, but this process is not included in our simulations since the model does not include a sub-grid turbulence parametrisation nor is at a suitable resolution to capture these processes. Future studies will need to consider the implications of turbulence, since this may have a significant effect on the cloud vertical structure. As cloud particles settle to deeper pressures, the total cloud cross-section reduces in the upper atmosphere, leading to a reduced opacity in the transmission region. While the timescales involved are typically too long to capture with current 3D simulations, it will be necessary in future studies to address this mechanism. The results from our enhanced precipitation tests, which simulate the effects of a more vertically evolved cloud deck, indicate that the ability to detect gaseous absorption signatures through the grey opacity, is strongly dependent on the cloud top pressure. The retrieval analysis of \citet{barstow17} on the \citet{sing16} dataset have previously indicated the importance of this connection, stressing the potential wide range of cloud top pressures across their small sample of hot-Jupiters. While we can, for the cloud structural and compositional conditions determined in \citet{lines18a}, constrain the cloud top pressure to be P $>$ 15 mbar for HD~209458b, the motivation of this result is clearly to more accurately pin-point the vertical equilibrium of the cloud (and hence upper boundary of the cloud opacity). It is also worth considering and cautioning that by parametrising the cloud top, the complex feedback between the cloud radiative transfer and the atmosphere's thermal, chemical and dynamical properties is circumvented. Thus, the resulting synthetic observations do not necessarily represent an atmosphere with a PT profile that has converged to the cloud vertical extent.


Observations of HD~209458b have revealed, repeatedly, the presence of the sodium \citep{charbonneau02,sing08,snellen08,jensen11} as well as water vapour and carbon monoxide \citep{snellen10,deming13}. It is interesting therefore to consider that while the literature typically considers HD~209458b to have a cloudless atmosphere due to the prominence of the aforementioned gaseous chemical signatures, our \citet{lines18a} simulations are remarkably cloudy in terms of the opacity and both horizontal and vertical distribution of cloud particles. As a result, our transmission spectrum probes that of an optically thick atmosphere, with many of the radiative interactions occurring within the first few vertical layers of dense silicate particles; cloud is radiatively dominant in the most upper layer. The result is an inevitably featureless profile, since there is no layer of gas absorbers above the cloud that can imprint on our spectra. Our opacity scaling tests indicate that a reduction in both the absorption and scattering coefficients of at least three orders of magnitude is required to see CO and water bands in the model spectrum, although CH$_4$ is visible in the full cloud opacity spectrum. We note that even with a strong reduction in the opacity, we are still unable to unmask the alkali metal spectral signatures. Our inability to emulate the observed spectrum of HD~209458b is an indication that we are missing physical processes or chemical constituents in our model\footnote{Caution must always be taken when making comparisons between transmission spectra from theoretical models and observations due to the presence of degeneracies, for example, those existing between baseline pressure, planetary radius and absorber abundances \citep[see][for more information]{benneke12,heng17b}.}. 

Our current model setup considers only five condensate materials and therefore may underestimate the atmospheric `cloudiness' by mass. The inclusion of a wider variety of important condensing species may lead to increased cloud particle sizes which will modify the cloud particle optical properties; the significance of this consideration is posited by the 1D microphysical study of \citet{powell18} who use a mass-binning technique for their cloud particles to reveal an irregular distribution of silicate particles and explore its importance. The current omission of iron itself, as a strong absorber, means we are likely underestimating cloud warming and this could lead to a situation whereby the atmospheric temperature increases, instead of the net cooling mechanism from a silicate dominant one. Aside from the aforementioned increase in condensate species, particle growth via coagulation could also play a role in producing heavier particles. Coagulation has been shown to be a requirement to drive precipitation in Earth's atmosphere \citep{pruppacher78}, with condensational growth alone not able to provide particle sizes large enough to initiate rainout. However, \citet{helling08-b1} shows that for sub-micron silicate grains in substellar atmospheres, coagulation can operate on a timescale orders of magnitude longer than chemical growth and therefore may not be an essential model component.

Our chosen initial metallicity may also play a significant role in cloud abundance, with non--solar values either reducing or increasing the total condensate mass and giving rise to a weakened or increased cloud opacity. This has been demonstrated in \citet{helling17b} for varying C/O ratios, and in \citet{mahapatra17} for rocky versus solar element abundances. Whatever the precise mechanism behind condensate growth, larger particles will effectively sediment out of the atmosphere and support the importance of investigating the role of precipitation.

We also acknowledge the omission of hydrocarbons formed photochemically in the gas--phase (e.g., Polycyclic Aromatic Hydrocarbons, or PAHs), which may play a large role in the atmospheric opacity and hence spectrum. Even if PAHs cannot themselves survive the large UV flux on the day side, their precursors radicals have been found to be quite abundant in the upper atmospheric layers \citep{venot15}. That situation may change as the radical--rich gas is advected by zonal winds from the hot/irradiated day side to the cold/dark night side, leading to the efficient formation of PAHs and, subsequently, photochemical hazes. It has been demonstrated however, that the abundances of PAHs can be very low, despite their abundances increasing in cloud forming regions due to the reduction of oxygen \citep{bilger13}.

These few physically motivated model adjustments alone may be enough to better reproduce the observed data, by way of increasing the detectability of gas--phase atomic and molecular absorption features. The benefit and power of using this physically motivated model is the ability to isolate and identify such specific physical processes and details which can be difficult for parametrised models which may obscure the underlying physical mechanisms at play.Therefore, although unlikely to represent the current conditions on HD~209458b, our simulations do provide an insight into those atmospheres which contain optically thick cloud, potentially from suspended silicate condensates. For example, our results are consistent with those of WASP-101b, which has a flat spectrum (there is no WFC3 H$_2$O feature) despite the planet possessing similar atmospheric properties to HD~209458b \citep{wakeford17b}. This is a possible indication that subtle differences in the atmospheric circulation, metallicity, C/O ratio etc. and interplay with the cloud radiative feedback could lead to large changes in the cloud's ability to impact on the observations. Additionally, despite the grey cloud opacity for our simulations, we are still able to identify transmission features which indicate cloud coverage and which have the potential for detection with JWST. Finally, we acknowledge the convenient ability to obtain synthetic transmission spectra directly from our 3D simulations which will enhance our ability to connect with the latest observations.

\section*{Acknowledgements}
SL and JG are funded by and thankful to the Leverhulme Trust. NJM is part funded by a Leverhulme Trust Research Project Grant. JM and IAB acknowledge the support of a Met Office Academic Partnership secondment. BD acknowledges funding from the European Research Council (ERC) under the European Unions Seventh Framework Programme (FP7/2007-2013) / ERC grant agreement no. 336792. GKHL acknowledges support from the Universities of Oxford and Bern through the Bernoulli fellowship program. A. L. C is funded by an STFC studentship. The calculations for this paper were performed on the University of Exeter Supercomputer, a DiRAC Facility jointly funded by STFC, the Large Facilities Capital Fund of BIS, and the University of Exeter. Material produced using Met Office Software. The authors would like to thank the referee for their insightful comments and suggestions.


\bibliographystyle{mnras}
\bibliography{clouds}

\begin{thebibliography}{}
\makeatletter
\relax
\def\mn@urlcharsother{\let\do\@makeother \do\$\do\&\do\#\do\^\do\_\do\%\do\~}
\def\mn@doi{\begingroup\mn@urlcharsother \@ifnextchar [ {\mn@doi@}
  {\mn@doi@[]}}
\def\mn@doi@[#1]#2{\def\@tempa{#1}\ifx\@tempa\@empty \href
  {http://dx.doi.org/#2} {doi:#2}\else \href {http://dx.doi.org/#2} {#1}\fi
  \endgroup}
\def\mn@eprint#1#2{\mn@eprint@#1:#2::\@nil}
\def\mn@eprint@arXiv#1{\href {http://arxiv.org/abs/#1} {{\tt arXiv:#1}}}
\def\mn@eprint@dblp#1{\href {http://dblp.uni-trier.de/rec/bibtex/#1.xml}
  {dblp:#1}}
\def\mn@eprint@#1:#2:#3:#4\@nil{\def\@tempa {#1}\def\@tempb {#2}\def\@tempc
  {#3}\ifx \@tempc \@empty \let \@tempc \@tempb \let \@tempb \@tempa \fi \ifx
  \@tempb \@empty \def\@tempb {arXiv}\fi \@ifundefined
  {mn@eprint@\@tempb}{\@tempb:\@tempc}{\expandafter \expandafter \csname
  mn@eprint@\@tempb\endcsname \expandafter{\@tempc}}}

\bibitem[\protect\citeauthoryear{{Amundsen}, {Baraffe}, {Tremblin}, {Manners},
  {Hayek}, {Mayne}  \& {Acreman}}{{Amundsen} et~al.}{2014}]{amundsen14}
{Amundsen} D.~S.,  {Baraffe} I.,  {Tremblin} P.,  {Manners} J.,  {Hayek} W.,
  {Mayne} N.~J.,   {Acreman} D.~M.,  2014, \aap, \href
  {http://adsabs.harvard.edu/abs/2014A%26A...564A..59A} {564, A59}

\bibitem[\protect\citeauthoryear{{Amundsen} et~al.,}{{Amundsen}
  et~al.}{2016}]{amundsen16}
{Amundsen} D.~S.,  et~al., 2016, \aap, \href
  {http://adsabs.harvard.edu/abs/2016A%26A...595A..36A} {595, A36}

\bibitem[\protect\citeauthoryear{{Amundsen}, {Tremblin}, {Manners}, {Baraffe}
  \& {Mayne}}{{Amundsen} et~al.}{2017}]{amundsen17}
{Amundsen} D.~S.,  {Tremblin} P.,  {Manners} J.,  {Baraffe} I.,   {Mayne}
  N.~J.,  2017, \aap, \href
  {http://adsabs.harvard.edu/abs/2017A%26A...598A..97A} {598, A97}

\bibitem[\protect\citeauthoryear{{Armstrong}, {de Mooij}, {Barstow}, {Osborn},
  {Blake}  \& {Saniee}}{{Armstrong} et~al.}{2016}]{armstrong16}
{Armstrong} D.~J.,  {de Mooij} E.,  {Barstow} J.,  {Osborn} H.~P.,  {Blake} J.,
    {Saniee} N.~F.,  2016, Nature Astronomy, \href
  {http://adsabs.harvard.edu/abs/2016NatAs...1E...4A} {1, 0004}

\bibitem[\protect\citeauthoryear{{Barstow}, {Aigrain}, {Irwin}  \&
  {Sing}}{{Barstow} et~al.}{2017}]{barstow17}
{Barstow} J.~K.,  {Aigrain} S.,  {Irwin} P.~G.~J.,   {Sing} D.~K.,  2017,
  \mn@doi [\apj] {10.3847/1538-4357/834/1/50}, \href
  {http://adsabs.harvard.edu/abs/2017ApJ...834...50B} {834, 50}

\bibitem[\protect\citeauthoryear{{Batalha} et~al.,}{{Batalha}
  et~al.}{2017}]{batalha17}
{Batalha} N.~E.,  et~al., 2017, \mn@doi [\pasp] {10.1088/1538-3873/aa65b0},
  \href {http://adsabs.harvard.edu/abs/2017PASP..129f4501B} {129, 064501}

\bibitem[\protect\citeauthoryear{{Benneke} \& {Seager}}{{Benneke} \&
  {Seager}}{2012}]{benneke12}
{Benneke} B.,  {Seager} S.,  2012, \mn@doi [\apj]
  {10.1088/0004-637X/753/2/100}, \href
  {http://adsabs.harvard.edu/abs/2012ApJ...753..100B} {753, 100}

\bibitem[\protect\citeauthoryear{{Bilger}, {Rimmer}  \& {Helling}}{{Bilger}
  et~al.}{2013}]{bilger13}
{Bilger} C.,  {Rimmer} P.,   {Helling} C.,  2013, \mn@doi [\mnras]
  {10.1093/mnras/stt1378}, \href
  {http://ukads.nottingham.ac.uk/abs/2013MNRAS.435.1888B} {435, 1888}

\bibitem[\protect\citeauthoryear{{Birkby}, {de Kok}, {Brogi}, {Schwarz}  \&
  {Snellen}}{{Birkby} et~al.}{2017}]{birkby17}
{Birkby} J.~L.,  {de Kok} R.~J.,  {Brogi} M.,  {Schwarz} H.,   {Snellen}
  I.~A.~G.,  2017, \aj, \href
  {http://adsabs.harvard.edu/abs/2017AJ....153..138B} {153, 138}

\bibitem[\protect\citeauthoryear{{Brogi}, {de Kok}, {Albrecht}, {Snellen},
  {Birkby}  \& {Schwarz}}{{Brogi} et~al.}{2016}]{brogi16}
{Brogi} M.,  {de Kok} R.~J.,  {Albrecht} S.,  {Snellen} I.~A.~G.,  {Birkby}
  J.~L.,   {Schwarz} H.,  2016, \apj, \href
  {http://adsabs.harvard.edu/abs/2016ApJ...817..106B} {817, 106}

\bibitem[\protect\citeauthoryear{{Brown}}{{Brown}}{2001}]{brown01}
{Brown} T.~M.,  2001, \mn@doi [\apj] {10.1086/320950}, \href
  {http://adsabs.harvard.edu/abs/2001ApJ...553.1006B} {553, 1006}

\bibitem[\protect\citeauthoryear{{Burrows}, {Rauscher}, {Spiegel}  \&
  {Menou}}{{Burrows} et~al.}{2010}]{burrows10}
{Burrows} A.,  {Rauscher} E.,  {Spiegel} D.~S.,   {Menou} K.,  2010, \mn@doi
  [\apj] {10.1088/0004-637X/719/1/341}, \href
  {http://adsabs.harvard.edu/abs/2010ApJ...719..341B} {719, 341}

\bibitem[\protect\citeauthoryear{{Carone}, {Keppens}  \& {Decin}}{{Carone}
  et~al.}{2015}]{carone15}
{Carone} L.,  {Keppens} R.,   {Decin} L.,  2015, \mn@doi [\mnras]
  {10.1093/mnras/stv1752}, \href
  {http://adsabs.harvard.edu/abs/2015MNRAS.453.2412C} {453, 2412}

\bibitem[\protect\citeauthoryear{{Carone}, {Keppens}  \& {Decin}}{{Carone}
  et~al.}{2016}]{carone16}
{Carone} L.,  {Keppens} R.,   {Decin} L.,  2016, \mn@doi [\mnras]
  {10.1093/mnras/stw1265}, \href
  {http://adsabs.harvard.edu/abs/2016MNRAS.461.1981C} {461, 1981}

\bibitem[\protect\citeauthoryear{{Charbonneau}, {Brown}, {Noyes}  \&
  {Gilliland}}{{Charbonneau} et~al.}{2002}]{charbonneau02}
{Charbonneau} D.,  {Brown} T.~M.,  {Noyes} R.~W.,   {Gilliland} R.~L.,  2002,
  \apj, \href {http://adsabs.harvard.edu/abs/2002ApJ...568..377C} {568, 377}

\bibitem[\protect\citeauthoryear{{Deming} et~al.,}{{Deming}
  et~al.}{2013}]{deming13}
{Deming} D.,  et~al., 2013, \apj, \href
  {http://adsabs.harvard.edu/abs/2013ApJ...774...95D} {774, 95}

\bibitem[\protect\citeauthoryear{{Demory} et~al.,}{{Demory}
  et~al.}{2013}]{demory13}
{Demory} B.-O.,  et~al., 2013, \apjl, \href
  {http://adsabs.harvard.edu/abs/2013ApJ...776L..25D} {776, L25}

\bibitem[\protect\citeauthoryear{{Dobbs-Dixon} \& {Agol}}{{Dobbs-Dixon} \&
  {Agol}}{2013}]{dobbs13}
{Dobbs-Dixon} I.,  {Agol} E.,  2013, \mnras, \href
  {http://adsabs.harvard.edu/abs/2013MNRAS.435.3159D} {435, 3159}

\bibitem[\protect\citeauthoryear{{Dobbs-Dixon}, {Agol}  \&
  {Burrows}}{{Dobbs-Dixon} et~al.}{2012}]{dobbs12}
{Dobbs-Dixon} I.,  {Agol} E.,   {Burrows} A.,  2012, \mn@doi [\apj]
  {10.1088/0004-637X/751/2/87}, \href
  {http://adsabs.harvard.edu/abs/2012ApJ...751...87D} {751, 87}

\bibitem[\protect\citeauthoryear{{Drummond} et~al.,}{{Drummond}
  et~al.}{2018}]{drummond18b}
{Drummond} B.,  et~al., 2018, \mn@doi [\apjl] {10.3847/2041-8213/aab209}, \href
  {http://adsabs.harvard.edu/abs/2018ApJ...855L..31D} {855, L31}

\bibitem[\protect\citeauthoryear{Edwards}{Edwards}{1996}]{edwards96b}
Edwards J.~M.,  1996, \mn@doi [Journal of the Atmospheric Sciences]
  {10.1175/1520-0469(1996)053<1921:ECOIFA>2.0.CO;2}, 53, 1921

\bibitem[\protect\citeauthoryear{Edwards \& Slingo}{Edwards \&
  Slingo}{1996}]{edwards96a}
Edwards J.~M.,  Slingo A.,  1996, \mn@doi [Quarterly Journal of the Royal
  Meteorological Society] {10.1002/qj.49712253107}, 122, 689

\bibitem[\protect\citeauthoryear{{Evans} et~al.,}{{Evans}
  et~al.}{2017}]{evans17}
{Evans} T.~M.,  et~al., 2017, \nat, \href
  {http://adsabs.harvard.edu/abs/2017Natur.548...58E} {548, 58}

\bibitem[\protect\citeauthoryear{{Fortney}, {Sudarsky}, {Hubeny}, {Cooper},
  {Hubbard}, {Burrows}  \& {Lunine}}{{Fortney} et~al.}{2003}]{fortney03}
{Fortney} J.~J.,  {Sudarsky} D.,  {Hubeny} I.,  {Cooper} C.~S.,  {Hubbard}
  W.~B.,  {Burrows} A.,   {Lunine} J.~I.,  2003, \mn@doi [\apj]
  {10.1086/374387}, \href {http://adsabs.harvard.edu/abs/2003ApJ...589..615F}
  {589, 615}

\bibitem[\protect\citeauthoryear{{Fortney}, {Shabram}, {Showman}, {Lian},
  {Freedman}, {Marley}  \& {Lewis}}{{Fortney} et~al.}{2010}]{fortney10}
{Fortney} J.~J.,  {Shabram} M.,  {Showman} A.~P.,  {Lian} Y.,  {Freedman}
  R.~S.,  {Marley} M.~S.,   {Lewis} N.~K.,  2010, \mn@doi [\apj]
  {10.1088/0004-637X/709/2/1396}, \href
  {http://adsabs.harvard.edu/abs/2010ApJ...709.1396F} {709, 1396}

\bibitem[\protect\citeauthoryear{{Goyal} et~al.,}{{Goyal}
  et~al.}{2018}]{goyal18}
{Goyal} J.~M.,  et~al., 2018, \mn@doi [\mnras] {10.1093/mnras/stx3015}, \href
  {http://adsabs.harvard.edu/abs/2018MNRAS.474.5158G} {474, 5158}

\bibitem[\protect\citeauthoryear{{Helling} \& {Woitke}}{{Helling} \&
  {Woitke}}{2006}]{helling06}
{Helling} C.,  {Woitke} P.,  2006, \aap, \href
  {http://adsabs.harvard.edu/abs/2006A%26A...455..325H} {455, 325}

\bibitem[\protect\citeauthoryear{{Helling}, {Thi}, {Woitke}  \&
  {Fridlund}}{{Helling} et~al.}{2006}]{helling06b}
{Helling} C.,  {Thi} W.-F.,  {Woitke} P.,   {Fridlund} M.,  2006, \mn@doi
  [\aap] {10.1051/0004-6361:20064944}, \href
  {http://ukads.nottingham.ac.uk/abs/2006A%26A...451L...9H} {451, L9}

\bibitem[\protect\citeauthoryear{{Helling}, {Woitke}  \& {Thi}}{{Helling}
  et~al.}{2008a}]{helling08}
{Helling} C.,  {Woitke} P.,   {Thi} W.-F.,  2008a, \aap, \href
  {http://adsabs.harvard.edu/abs/2008A%26A...485..547H} {485, 547}

\bibitem[\protect\citeauthoryear{{Helling}, {Dehn}, {Woitke}  \&
  {Hauschildt}}{{Helling} et~al.}{2008b}]{helling08-b1}
{Helling} C.,  {Dehn} M.,  {Woitke} P.,   {Hauschildt} P.~H.,  2008b, \apjl,
  \href {http://adsabs.harvard.edu/abs/2008ApJ...675L.105H} {675, L105}

\bibitem[\protect\citeauthoryear{{Helling} et~al.,}{{Helling}
  et~al.}{2016}]{helling16}
{Helling} C.,  et~al., 2016, \mnras, \href
  {http://adsabs.harvard.edu/abs/2016MNRAS.460..855H} {460, 855}

\bibitem[\protect\citeauthoryear{{Helling}, {Tootill}, {Woitke}  \&
  {Lee}}{{Helling} et~al.}{2017}]{helling17b}
{Helling} C.,  {Tootill} D.,  {Woitke} P.,   {Lee} G.,  2017, \mn@doi [\aap]
  {10.1051/0004-6361/201629696}, \href
  {http://adsabs.harvard.edu/abs/2017A%26A...603A.123H} {603, A123}

\bibitem[\protect\citeauthoryear{{Heng} \& {Kitzmann}}{{Heng} \&
  {Kitzmann}}{2017}]{heng17b}
{Heng} K.,  {Kitzmann} D.,  2017, \mn@doi [\mnras] {10.1093/mnras/stx1453},
  \href {http://adsabs.harvard.edu/abs/2017MNRAS.470.2972H} {470, 2972}

\bibitem[\protect\citeauthoryear{{Heng} \& {Showman}}{{Heng} \&
  {Showman}}{2015}]{heng15}
{Heng} K.,  {Showman} A.~P.,  2015, Annual Review of Earth and Planetary
  Sciences, \href {http://adsabs.harvard.edu/abs/2015AREPS..43..509H} {43, 509}

\bibitem[\protect\citeauthoryear{{Heng}, {Menou}  \& {Phillipps}}{{Heng}
  et~al.}{2011}]{heng11}
{Heng} K.,  {Menou} K.,   {Phillipps} P.~J.,  2011, \mnras, \href
  {http://adsabs.harvard.edu/abs/2011MNRAS.413.2380H} {413, 2380}

\bibitem[\protect\citeauthoryear{{Hubbard}, {Fortney}, {Lunine}, {Burrows},
  {Sudarsky}  \& {Pinto}}{{Hubbard} et~al.}{2001}]{hubbard01}
{Hubbard} W.~B.,  {Fortney} J.~J.,  {Lunine} J.~I.,  {Burrows} A.,  {Sudarsky}
  D.,   {Pinto} P.,  2001, \mn@doi [\apj] {10.1086/322490}, \href
  {http://adsabs.harvard.edu/abs/2001ApJ...560..413H} {560, 413}

\bibitem[\protect\citeauthoryear{Iyer, Swain, Zellem, Line, Roudier, Rocha  \&
  Livingston}{Iyer et~al.}{2016}]{iyer16}
Iyer A.~R.,  Swain M.~R.,  Zellem R.~T.,  Line M.~R.,  Roudier G.,  Rocha G.,
  Livingston J.~H.,  2016, The Astrophysical Journal, 823, 109

\bibitem[\protect\citeauthoryear{{Jensen}, {Redfield}, {Endl}, {Cochran},
  {Koesterke}  \& {Barman}}{{Jensen} et~al.}{2011}]{jensen11}
{Jensen} A.~G.,  {Redfield} S.,  {Endl} M.,  {Cochran} W.~D.,  {Koesterke} L.,
   {Barman} T.~S.,  2011, \mn@doi [\apj] {10.1088/0004-637X/743/2/203}, \href
  {http://adsabs.harvard.edu/abs/2011ApJ...743..203J} {743, 203}

\bibitem[\protect\citeauthoryear{{Kataria}, {Sing}, {Lewis}, {Visscher},
  {Showman}, {Fortney}  \& {Marley}}{{Kataria} et~al.}{2016}]{kataria16}
{Kataria} T.,  {Sing} D.~K.,  {Lewis} N.~K.,  {Visscher} C.,  {Showman} A.~P.,
  {Fortney} J.~J.,   {Marley} M.~S.,  2016, \mn@doi [\apj]
  {10.3847/0004-637X/821/1/9}, \href
  {http://adsabs.harvard.edu/abs/2016ApJ...821....9K} {821, 9}

\bibitem[\protect\citeauthoryear{{Kirk}, {Wheatley}, {Louden}, {Doyle},
  {Skillen}, {McCormac}, {Irwin}  \& {Karjalainen}}{{Kirk}
  et~al.}{2017}]{kirk17}
{Kirk} J.,  {Wheatley} P.~J.,  {Louden} T.,  {Doyle} A.~P.,  {Skillen} I.,
  {McCormac} J.,  {Irwin} P.~G.~J.,   {Karjalainen} R.,  2017, \mnras, \href
  {http://adsabs.harvard.edu/abs/2017MNRAS.468.3907K} {468, 3907}

\bibitem[\protect\citeauthoryear{{Kitzmann} \& {Heng}}{{Kitzmann} \&
  {Heng}}{2018}]{kitzmann18}
{Kitzmann} D.,  {Heng} K.,  2018, \mn@doi [\mnras] {10.1093/mnras/stx3141},
  \href {http://adsabs.harvard.edu/abs/2018MNRAS.475...94K} {475, 94}

\bibitem[\protect\citeauthoryear{{Knutson}, {Charbonneau}, {Burrows},
  {O'Donovan}  \& {Mandushev}}{{Knutson} et~al.}{2009}]{knutson08}
{Knutson} H.~A.,  {Charbonneau} D.,  {Burrows} A.,  {O'Donovan} F.~T.,
  {Mandushev} G.,  2009, \mn@doi [\apj] {10.1088/0004-637X/691/1/866}, \href
  {http://adsabs.harvard.edu/abs/2009ApJ...691..866K} {691, 866}

\bibitem[\protect\citeauthoryear{{Lecavelier Des Etangs}, {Pont},
  {Vidal-Madjar}  \& {Sing}}{{Lecavelier Des Etangs} et~al.}{2008}]{etangs08}
{Lecavelier Des Etangs} A.,  {Pont} F.,  {Vidal-Madjar} A.,   {Sing} D.,  2008,
  \aap, \href {http://adsabs.harvard.edu/abs/2008A%26A...481L..83L} {481, L83}

\bibitem[\protect\citeauthoryear{{Lee}, {Irwin}, {Fletcher}, {Heng}  \&
  {Barstow}}{{Lee} et~al.}{2014}]{lee14}
{Lee} J.-M.,  {Irwin} P.~G.~J.,  {Fletcher} L.~N.,  {Heng} K.,   {Barstow}
  J.~K.,  2014, \mn@doi [\apj] {10.1088/0004-637X/789/1/14}, \href
  {http://adsabs.harvard.edu/abs/2014ApJ...789...14L} {789, 14}

\bibitem[\protect\citeauthoryear{{Lee}, {Helling}, {Dobbs-Dixon}  \&
  {Juncher}}{{Lee} et~al.}{2015}]{lee15a}
{Lee} G.,  {Helling} C.,  {Dobbs-Dixon} I.,   {Juncher} D.,  2015, \aap, \href
  {http://adsabs.harvard.edu/abs/2015A%26A...580A..12L} {580, A12}

\bibitem[\protect\citeauthoryear{{Lee}, {Dobbs-Dixon}, {Helling}, {Bognar}  \&
  {Woitke}}{{Lee} et~al.}{2016}]{lee16}
{Lee} G.,  {Dobbs-Dixon} I.,  {Helling} C.,  {Bognar} K.,   {Woitke} P.,  2016,
  \aap, \href {http://adsabs.harvard.edu/abs/2016A%26A...594A..48L} {594, A48}

\bibitem[\protect\citeauthoryear{{Line} et~al.,}{{Line} et~al.}{2016}]{line16}
{Line} M.~R.,  et~al., 2016, \mn@doi [\aj] {10.3847/0004-6256/152/6/203}, \href
  {http://adsabs.harvard.edu/abs/2016AJ....152..203L} {152, 203}

\bibitem[\protect\citeauthoryear{{Lines} et~al.,}{{Lines}
  et~al.}{2018}]{lines18a}
{Lines} S.,  et~al., 2018, \mn@doi [\aap] {10.1051/0004-6361/201732278}, \href
  {http://adsabs.harvard.edu/abs/2018A%26A...615A..97L} {615, A97}

\bibitem[\protect\citeauthoryear{{Louden} \& {Wheatley}}{{Louden} \&
  {Wheatley}}{2015}]{louden15}
{Louden} T.,  {Wheatley} P.~J.,  2015, \apjl, \href
  {http://adsabs.harvard.edu/abs/2015ApJ...814L..24L} {814, L24}

\bibitem[\protect\citeauthoryear{{Mahapatra}, {Helling}  \&
  {Miguel}}{{Mahapatra} et~al.}{2017}]{mahapatra17}
{Mahapatra} G.,  {Helling} C.,   {Miguel} Y.,  2017, \mn@doi [\mnras]
  {10.1093/mnras/stx1666}, \href
  {http://adsabs.harvard.edu/abs/2017MNRAS.472..447M} {472, 447}

\bibitem[\protect\citeauthoryear{{Marley} \& {Robinson}}{{Marley} \&
  {Robinson}}{2015}]{marley15}
{Marley} M.~S.,  {Robinson} T.~D.,  2015, \mn@doi [\araa]
  {10.1146/annurev-astro-082214-122522}, \href
  {http://ukads.nottingham.ac.uk/abs/2015ARA%26A..53..279M} {53, 279}

\bibitem[\protect\citeauthoryear{{Marley}, {Ackerman}, {Cuzzi}  \&
  {Kitzmann}}{{Marley} et~al.}{2013}]{marley13}
{Marley} M.~S.,  {Ackerman} A.~S.,  {Cuzzi} J.~N.,   {Kitzmann} D.,  2013,
  {Clouds and Hazes in Exoplanet Atmospheres}.
UAPress, p.~367, \mn@doi{10.2458/azu_uapress_9780816530595-ch15}

\bibitem[\protect\citeauthoryear{{Mayne}, {Baraffe}, {Acreman}, {Smith},
  {Wood}, {Amundsen}, {Thuburn}  \& {Jackson}}{{Mayne} et~al.}{2014a}]{mayne13}
{Mayne} N.~J.,  {Baraffe} I.,  {Acreman} D.~M.,  {Smith} C.,  {Wood} N.,
  {Amundsen} D.~S.,  {Thuburn} J.,   {Jackson} D.~R.,  2014a, Geoscientific
  Model Development, \href {http://adsabs.harvard.edu/abs/2014GMD.....7.3059M}
  {7, 3059}

\bibitem[\protect\citeauthoryear{{Mayne} et~al.,}{{Mayne}
  et~al.}{2014b}]{mayne14}
{Mayne} N.~J.,  et~al., 2014b, A$\&$A, \href
  {http://adsabs.harvard.edu/abs/2014A%26A...561A...1M} {561, A1}

\bibitem[\protect\citeauthoryear{{Mayne} et~al.,}{{Mayne}
  et~al.}{2017}]{mayne17}
{Mayne} N.~J.,  et~al., 2017, \aap, \href
  {http://adsabs.harvard.edu/abs/2017A%26A...604A..79M} {604, A79}

\bibitem[\protect\citeauthoryear{{Menou} \& {Rauscher}}{{Menou} \&
  {Rauscher}}{2009}]{menou09}
{Menou} K.,  {Rauscher} E.,  2009, \apj, \href
  {http://adsabs.harvard.edu/abs/2009ApJ...700..887M} {700, 887}

\bibitem[\protect\citeauthoryear{{Miller-Ricci Kempton} \&
  {Rauscher}}{{Miller-Ricci Kempton} \& {Rauscher}}{2012}]{kempton12}
{Miller-Ricci Kempton} E.,  {Rauscher} E.,  2012, \mn@doi [\apj]
  {10.1088/0004-637X/751/2/117}, \href
  {http://adsabs.harvard.edu/abs/2012ApJ...751..117M} {751, 117}

\bibitem[\protect\citeauthoryear{{Min}, {Hovenier}, {Waters}  \& {de
  Koter}}{{Min} et~al.}{2008}]{min08}
{Min} M.,  {Hovenier} J.~W.,  {Waters} L.~B.~F.~M.,   {de Koter} A.,  2008,
  \mn@doi [\aap] {10.1051/0004-6361:200809534}, \href
  {http://ukads.nottingham.ac.uk/abs/2008A%26A...489..135M} {489, 135}

\bibitem[\protect\citeauthoryear{{Nikolov} et~al.,}{{Nikolov}
  et~al.}{2015}]{nikolov15}
{Nikolov} N.,  et~al., 2015, \mn@doi [\mnras] {10.1093/mnras/stu2433}, \href
  {http://adsabs.harvard.edu/abs/2015MNRAS.447..463N} {447, 463}

\bibitem[\protect\citeauthoryear{{Oreshenko}, {Heng}  \& {Demory}}{{Oreshenko}
  et~al.}{2016}]{oreshenko16}
{Oreshenko} M.,  {Heng} K.,   {Demory} B.-O.,  2016, \mn@doi [\mnras]
  {10.1093/mnras/stw133}, \href
  {http://adsabs.harvard.edu/abs/2016MNRAS.457.3420O} {457, 3420}

\bibitem[\protect\citeauthoryear{{Parmentier}, {Showman}  \&
  {Lian}}{{Parmentier} et~al.}{2013}]{parmentier13}
{Parmentier} V.,  {Showman} A.~P.,   {Lian} Y.,  2013, \aap, \href
  {http://adsabs.harvard.edu/abs/2013A%26A...558A..91P} {558, A91}

\bibitem[\protect\citeauthoryear{{Parmentier}, {Fortney}, {Showman}, {Morley}
  \& {Marley}}{{Parmentier} et~al.}{2016}]{parmentier16}
{Parmentier} V.,  {Fortney} J.~J.,  {Showman} A.~P.,  {Morley} C.,   {Marley}
  M.~S.,  2016, \apj, \href {http://adsabs.harvard.edu/abs/2016ApJ...828...22P}
  {828, 22}

\bibitem[\protect\citeauthoryear{{Pont}, {Knutson}, {Gilliland}, {Moutou}  \&
  {Charbonneau}}{{Pont} et~al.}{2008}]{pont08}
{Pont} F.,  {Knutson} H.,  {Gilliland} R.~L.,  {Moutou} C.,   {Charbonneau} D.,
   2008, \mn@doi [\mnras] {10.1111/j.1365-2966.2008.12852.x}, \href
  {http://adsabs.harvard.edu/abs/2008MNRAS.385..109P} {385, 109}

\bibitem[\protect\citeauthoryear{{Powell}, {Zhang}, {Gao}  \&
  {Parmentier}}{{Powell} et~al.}{2018}]{powell18}
{Powell} D.,  {Zhang} X.,  {Gao} P.,   {Parmentier} V.,  2018, \mn@doi [\apj]
  {10.3847/1538-4357/aac215}, \href
  {http://adsabs.harvard.edu/abs/2018ApJ...860...18P} {860, 18}

\bibitem[\protect\citeauthoryear{{Pruppacher} \& {Klett}}{{Pruppacher} \&
  {Klett}}{1978}]{pruppacher78}
{Pruppacher} H.~R.,  {Klett} J.~D.,  1978, Microphysics of clouds and
  precipitation.
D. Reidel Pub. Co Dordrecht, Holland ; Boston

\bibitem[\protect\citeauthoryear{{Rauscher} \& {Kempton}}{{Rauscher} \&
  {Kempton}}{2014}]{rauscher14}
{Rauscher} E.,  {Kempton} E.~M.~R.,  2014, \mn@doi [\apj]
  {10.1088/0004-637X/790/1/79}, \href
  {http://adsabs.harvard.edu/abs/2014ApJ...790...79R} {790, 79}

\bibitem[\protect\citeauthoryear{{Rauscher} \& {Menou}}{{Rauscher} \&
  {Menou}}{2013}]{rauscher13}
{Rauscher} E.,  {Menou} K.,  2013, \mn@doi [\apj]
  {10.1088/0004-637X/764/1/103}, \href
  {http://adsabs.harvard.edu/abs/2013ApJ...764..103R} {764, 103}

\bibitem[\protect\citeauthoryear{{Roman} \& {Rauscher}}{{Roman} \&
  {Rauscher}}{2018}]{roman18}
{Roman} M.,  {Rauscher} E.,  2018, preprint, \href
  {http://adsabs.harvard.edu/abs/2018arXiv180708890R} {} (\mn@eprint {arXiv}
  {1807.08890})

\bibitem[\protect\citeauthoryear{{Seager} \& {Sasselov}}{{Seager} \&
  {Sasselov}}{2000}]{seager00}
{Seager} S.,  {Sasselov} D.~D.,  2000, \mn@doi [\apj] {10.1086/309088}, \href
  {http://adsabs.harvard.edu/abs/2000ApJ...537..916S} {537, 916}

\bibitem[\protect\citeauthoryear{{Showman} \& {Guillot}}{{Showman} \&
  {Guillot}}{2002}]{showman02}
{Showman} A.~P.,  {Guillot} T.,  2002, \aap, \href
  {http://adsabs.harvard.edu/abs/2002A%26A...385..166S} {385, 166}

\bibitem[\protect\citeauthoryear{{Showman} \& {Polvani}}{{Showman} \&
  {Polvani}}{2011}]{showman11}
{Showman} A.~P.,  {Polvani} L.~M.,  2011, \apj, \href
  {http://adsabs.harvard.edu/abs/2011ApJ...738...71S} {738, 71}

\bibitem[\protect\citeauthoryear{Showman, Fortney, Lian, Marley, Freedman,
  Knutson  \& Charbonneau}{Showman et~al.}{2009}]{showman09}
Showman A.~P.,  Fortney J.~J.,  Lian Y.,  Marley M.~S.,  Freedman R.~S.,
  Knutson H.~A.,   Charbonneau D.,  2009, The Astrophysical Journal, 699, 564

\bibitem[\protect\citeauthoryear{{Showman}, {Fortney}, {Lewis}  \&
  {Shabram}}{{Showman} et~al.}{2013}]{showman13}
{Showman} A.~P.,  {Fortney} J.~J.,  {Lewis} N.~K.,   {Shabram} M.,  2013,
  \mn@doi [\apj] {10.1088/0004-637X/762/1/24}, \href
  {http://adsabs.harvard.edu/abs/2013ApJ...762...24S} {762, 24}

\bibitem[\protect\citeauthoryear{{Sing}, {Vidal-Madjar}, {D{\'e}sert},
  {Lecavelier des Etangs}  \& {Ballester}}{{Sing} et~al.}{2008}]{sing08}
{Sing} D.~K.,  {Vidal-Madjar} A.,  {D{\'e}sert} J.-M.,  {Lecavelier des Etangs}
  A.,   {Ballester} G.,  2008, \mn@doi [\apj] {10.1086/590075}, \href
  {http://adsabs.harvard.edu/abs/2008ApJ...686..658S} {686, 658}

\bibitem[\protect\citeauthoryear{{Sing} et~al.,}{{Sing} et~al.}{2011}]{sing11}
{Sing} D.~K.,  et~al., 2011, \aap, \href
  {http://adsabs.harvard.edu/abs/2011A%26A...527A..73S} {527, A73}

\bibitem[\protect\citeauthoryear{{Sing} et~al.,}{{Sing} et~al.}{2016}]{sing16}
{Sing} D.~K.,  et~al., 2016, Nature, \href
  {http://adsabs.harvard.edu/abs/2016Natur.529...59S} {529, 59}

\bibitem[\protect\citeauthoryear{{Snellen}, {Albrecht}, {de Mooij}  \& {Le
  Poole}}{{Snellen} et~al.}{2008}]{snellen08}
{Snellen} I.~A.~G.,  {Albrecht} S.,  {de Mooij} E.~J.~W.,   {Le Poole} R.~S.,
  2008, \mn@doi [\aap] {10.1051/0004-6361:200809762}, \href
  {http://adsabs.harvard.edu/abs/2008A%26A...487..357S} {487, 357}

\bibitem[\protect\citeauthoryear{{Snellen}, {de Kok}, {de Mooij}  \&
  {Albrecht}}{{Snellen} et~al.}{2010}]{snellen10}
{Snellen} I.~A.~G.,  {de Kok} R.~J.,  {de Mooij} E.~J.~W.,   {Albrecht} S.,
  2010, \mn@doi [\nat] {10.1038/nature09111}, \href
  {http://adsabs.harvard.edu/abs/2010Natur.465.1049S} {465, 1049}

\bibitem[\protect\citeauthoryear{{Stevenson} et~al.,}{{Stevenson}
  et~al.}{2014}]{stevenson14}
{Stevenson} K.~B.,  et~al., 2014, \mn@doi [Science] {10.1126/science.1256758},
  \href {http://adsabs.harvard.edu/abs/2014Sci...346..838S} {346, 838}

\bibitem[\protect\citeauthoryear{{Stevenson} et~al.,}{{Stevenson}
  et~al.}{2017}]{stevenson17}
{Stevenson} K.~B.,  et~al., 2017, \aj, \href
  {http://adsabs.harvard.edu/abs/2017AJ....153...68S} {153, 68}

\bibitem[\protect\citeauthoryear{{Tremblin} et~al.,}{{Tremblin}
  et~al.}{2017}]{tremblin17}
{Tremblin} P.,  et~al., 2017, \apj, \href
  {http://adsabs.harvard.edu/abs/2017ApJ...841...30T} {841, 30}

\bibitem[\protect\citeauthoryear{{Venot}, {H{\'e}brard}, {Ag{\'u}ndez}, {Decin}
   \& {Bounaceur}}{{Venot} et~al.}{2015}]{venot15}
{Venot} O.,  {H{\'e}brard} E.,  {Ag{\'u}ndez} M.,  {Decin} L.,   {Bounaceur}
  R.,  2015, \mn@doi [\aap] {10.1051/0004-6361/201425311}, \href
  {http://adsabs.harvard.edu/abs/2015A%26A...577A..33V} {577, A33}

\bibitem[\protect\citeauthoryear{{Wakeford} \& {Sing}}{{Wakeford} \&
  {Sing}}{2015}]{wakeford15}
{Wakeford} H.~R.,  {Sing} D.~K.,  2015, \mn@doi [\aap]
  {10.1051/0004-6361/201424207}, \href
  {http://adsabs.harvard.edu/abs/2015A%26A...573A.122W} {573, A122}

\bibitem[\protect\citeauthoryear{{Wakeford} et~al.,}{{Wakeford}
  et~al.}{2017}]{wakeford17b}
{Wakeford} H.~R.,  et~al., 2017, \mn@doi [\apjl] {10.3847/2041-8213/835/1/L12},
  \href {http://adsabs.harvard.edu/abs/2017ApJ...835L..12W} {835, L12}

\bibitem[\protect\citeauthoryear{{Woitke} \& {Helling}}{{Woitke} \&
  {Helling}}{2003}]{woitke03}
{Woitke} P.,  {Helling} C.,  2003, \aap, \href
  {http://adsabs.harvard.edu/abs/2003A%26A...399..297W} {399, 297}

\bibitem[\protect\citeauthoryear{{Woitke} \& {Helling}}{{Woitke} \&
  {Helling}}{2004}]{woitke04}
{Woitke} P.,  {Helling} C.,  2004, \aap, \href
  {http://adsabs.harvard.edu/abs/2004A%26A...414..335W} {414, 335}

\bibitem[\protect\citeauthoryear{{Y-K.~Cho}, {Menou}, {Hansen}  \&
  {Seager}}{{Y-K.~Cho} et~al.}{2006}]{cho06}
{Y-K.~Cho} J.,  {Menou} K.,  {Hansen} B.,   {Seager} S.,  2006, ArXiv
  Astrophysics e-prints, \href
  {http://adsabs.harvard.edu/abs/2006astro.ph..7338Y} {}

\bibitem[\protect\citeauthoryear{Zdunkowski, Welch  \& Korb}{Zdunkowski
  et~al.}{1980}]{zdunkowski80}
Zdunkowski W.,  Welch R.,   Korb G.,  1980, Beitr{\"a}ge zur Physik der
  Atmosph{\"a}re, 53, 147

\bibitem[\protect\citeauthoryear{{Zeidler}, {Posch}  \& {Mutschke}}{{Zeidler}
  et~al.}{2013}]{zeidler13}
{Zeidler} S.,  {Posch} T.,   {Mutschke} H.,  2013, \mn@doi [\aap]
  {10.1051/0004-6361/201220459}, \href
  {http://ukads.nottingham.ac.uk/abs/2013A%26A...553A..81Z} {553, A81}

\bibitem[\protect\citeauthoryear{{Zeidler}, {Mutschke}  \& {Posch}}{{Zeidler}
  et~al.}{2015}]{zeidler15}
{Zeidler} S.,  {Mutschke} H.,   {Posch} T.,  2015, \mn@doi [\apj]
  {10.1088/0004-637X/798/2/125}, \href
  {http://ukads.nottingham.ac.uk/abs/2015ApJ...798..125Z} {798, 125}

\bibitem[\protect\citeauthoryear{Zhang, Strobel  \& Imanaka}{Zhang
  et~al.}{2017}]{zhang17}
Zhang X.,  Strobel D.~F.,   Imanaka H.,  2017, Nature, 551, 352

\makeatother
\end{thebibliography}

\appendix
\section{Appendix}


\begin{figure*}
\centering
\includegraphics[scale=0.85]{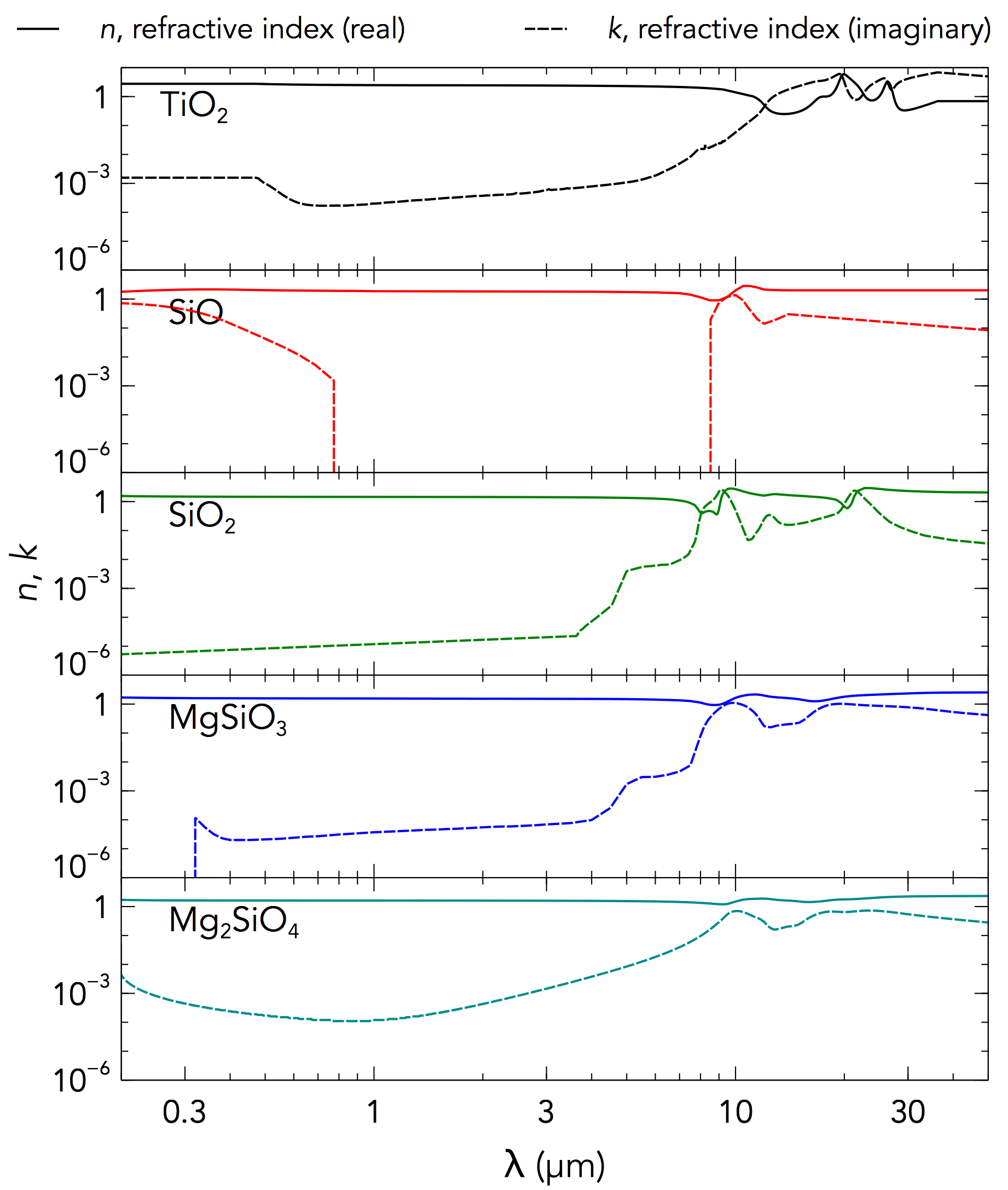}
\caption{Complex refractive index for the 5 dust species that make up our mixed-composition mineral cloud particles. The real component of refractive index, $n$, is shown in solid lines and the imaginary component, or extinction coefficient $k$, as dashed lines. A reference list of sources for the optical constants can be found in \citet{lee15a}.}
\label{fig:nk}
\end{figure*}


\end{document}